\documentclass[%
 reprint,
nofootinbib,
 amsmath,amssymb,
 aps,
]{revtex4-2}

\usepackage[normalem]{ulem}
\usepackage{comment}
\usepackage[super]{nth}
\usepackage{graphicx}
\usepackage{dcolumn}
\usepackage{bm}
\usepackage[dvipsnames]{xcolor}
\usepackage{pgfplotstable, booktabs}
\usepackage{array}

\usepackage{hyperref}

\begin{document}

\preprint{APS/123-QED}

\title{High-energy interactions of charged black holes in full general relativity II: Near-extremal merger remnants and universality with the irreducible mass}

\author{M A. M. Smith} 
\email{mamsmith@arizona.edu, they/them/theirs}
\affiliation{%
 Department of Physics, The University of Arizona, Tucson, AZ, USA
}%
\author{Vasileios Paschalidis}
\email{vpaschal@arizona.edu}
\affiliation{Department of Astronomy, The University of Arizona, Tucson, AZ, USA
}
\affiliation{Department of Physics, The University of Arizona, Tucson, AZ, USA}

\author{Gabriele Bozzola}
\affiliation{Division of Geological and Planetary Sciences, California Institute of Technology, Pasadena, CA, USA}
\affiliation{Department of Astronomy, The University of Arizona, Tucson, AZ, USA}

\date{\today}

\begin{abstract}
In a previous paper~[\href{https://arxiv.org/abs/2411.11960}{arXiv:2411.11960 [gr-qc] (2024)}], we initiated a study of high-energy interactions of charged binary black holes near the scattering threshold, focusing on zoom-whirl orbits. In this second paper in our series, we focus on merger remnant properties and energetics with new simulations of equal-mass, equal-charge, nonspinning binary black holes with variable impact parameter. We find near-extremal merger remnants with Kerr-Newman parameter reaching  $\Upsilon_f = 0.97$, and observe that the maximum $\Upsilon_f$ increases monotonically with $\lambda$ for a fixed initial Lorentz factor. We find that binaries with larger $\lambda$ radiate less total energy despite having stronger electromagnetic emission. The maximum energy radiated by a binary in our study is $31\%$ of its gravitational mass. Increasing $\lambda$ has little effect on the maximum angular momentum radiated, which was $\approx 72\%$ of the spacetime total angular momentum for each $\lambda$ explored here. Lastly, we provide additional evidence for the universality with the irreducible mass that we discovered in~[\href{https://arxiv.org/abs/2411.11960}{arXiv:2411.11960 [gr-qc] (2024)}]. The black hole horizon areal radius determines a fundamental, gauge-invariant length scale governing BH interactions near the scattering threshold. 
\end{abstract}

\maketitle


\section{Introduction}

We continue our investigation of high-energy charged black hole (BH) collisions that we initiated in~\cite{smith_high-energy_2024} (Paper I henceforth). Our work is motivated by non-linear effects that have previously been observed near the scattering threshold regime for uncharged binary black holes (BBHs): 1) zoom-whirl orbits~\cite{bombelli_chaos_1992,levin_gravity_2000,glampedakis_zoom_2002,pretorius_black_2007,healy_zoom-whirl_2009,sperhake_cross_2009,gold_radiation_2010,gundlach_critical_2012}, 2) the production of near-extremal merger remnants~\cite{sperhake_cross_2009}, and 3) intense amounts of radiated energy and angular momentum~\cite{sperhake_cross_2009,shibata_high-velocity_2008, sperhake_universality_2013}. In these previous studies it was found that merger remnants with near-extremal dimensionless spins, \(j_f > 0.95\), can be produced from initially nonspinning BHs~\cite{sperhake_cross_2009} near the scattering threshold, and $j_f>0.9865$ with initial dimensionless spin up to 0.8~\cite{healy_maximum_2024}. Uncharged, initially nonspinning  BBHs can also radiate \(\sim30\%\) of their initial energy (or $\sim40\%$ for larger initial boosts
~\cite{sperhake_universality_2013}) and \(\sim 70\%\) of their initial angular momentum \cite{sperhake_cross_2009,shibata_high-velocity_2008, sperhake_universality_2013}, making high-energy encounters of BHs among the most energetic events. These results exemplify processes that probe general relativity to the extreme, thereby enabling a deeper understanding of the theory. Moreover, extreme scattering processes involving BHs can be used to investigate cosmic censorship, the maximum luminosity of physical processes, the effect of internal BH structure, trans-Planckian scattering, and AdS/CFT correspondence to name a few (see, e.g.,~\cite{cardoso_exploring_2015}).

The highly non-linear scattering regime is determined by a threshold impact parameter $b_{scat}$, which separates binaries that merge ($b < b_{scat}$) from binaries that scatter back out to infinity ($b > b_{scat}$). An additional threshold impact parameter exists, $b^* < b_{scat}$, that separates binaries that merge immediately (in a single encounter, $b<b^*$) from those that merge after multiple close encounters ($b^*< b < b_{scat}$) --- the zoom-whirl regime (see Paper I for more details).

In Paper I we investigated the effect of U(1) charge on zoom-whirl orbits. We simulated the scattering of equal-mass, equal-charge (with initial charge-to-mass ratios $\lambda \leq 0.6$), nonspinning binary black holes (BBH) with initial Lorentz factor $\gamma\simeq 1.52$ and impact parameters near the scattering and immediate merger thresholds. We found binaries that exhibited zoom-whirl orbits for all values of $\lambda$ we probed. We observed that the values of the immediate merger and scattering threshold impact parameters decreased with increasing $\lambda$ when normalized by the Arnowitt-Deser-Misner (ADM) mass of the spacetime ($M_{\rm ADM}$). A key result in our study was that the immediate merger and scattering thresholds became universal --- charge-independent --- when normalized by the sum of the initial BH horizon areal radii, which is proportional to the sum of their irreducible masses ($M_{\rm irr}$). In this work we continue our investigation of extreme collisions of charged BHs by
exploring this universality further, and by focusing on the energetics of these interactions, as well as the properties of the remnants that form for $b<b_{\rm scat}$. 

There are a number of unanswered questions about extreme interactions of charged black holes: First, how is the extremality of merger remnants in this regime affected by charge? In uncharged binaries, the extremality depends on the ratio of the remnant spin angular momentum ($J$) to its gravitational mass (${\cal M}$) squared. Both of these quantities can be radiated away, unlike charge. For charged BHs, extremality is described by the Kerr-Newman parameter $\Upsilon$~\cite{newman_metric_1965}~\footnote{For black holes, the Kerr-Newman parameter satisfies $\Upsilon \leq 1$.}, 
\begin{eqnarray}\label{eq:chif_def}
\Upsilon =\sqrt{\frac{J^2}{{\cal M}^4} + \frac{Q^2}{{\cal M}^2}}
\label{eq:chi_formula},
\end{eqnarray}
where \(Q\) is the BH charge. How do the BBH angular momentum, mass, and charge interplay to affect the extremality of the merger remnant BH when the first two are not conserved, but the third is conserved? Are even more extremal remnants produced when the initial charge-to-mass ratio of the BHs is increased and the impact parameter held constant?  What is the most extremal remnant that can arise when charge is included? This leads to questions of cosmic censorship: could \(\Upsilon_f\) (the subscript $f$ denotes properties of a merger remnant throughout this work) exceed unity, implying the formation of a naked singularity?  Note that no pathway to violating cosmic censorship was found for BH head-on collisions when accounting for charge~\cite{zilhao_collisions_2012, bozzola_does_2022} or quasi-circular mergers of nonspinning black holes~\cite{bozzola_can_2023}. However, BBH configurations near the scattering threshold can radiate significant amounts energy and angular momentum, thereby opening a new path to testing cosmic censorship.

Second, how does charge affect the energy and angular momentum radiated as the impact parameter approaches \(b_{\rm scat}\)? Charged BHs emit electromagnetic radiation in addition to gravitational radiation; can previous limits on the maximum radiated energy or angular momentum be surpassed when charge is accounted for~\cite{shibata_high-velocity_2008,sperhake_cross_2009,sperhake_universality_2013}? 

Finally, in Paper I we observed that, while the mantra ``matter doesn't matter" \cite{hooft_graviton_1987,amati_superstring_1987,banks_model_1999,choptuik_ur_2010, rezzolla_black_2012,east_ultrarelativistic_2013,sperhake_unequal_2016} holds in the case of head-on collisions even at moderate Lorentz factors~\cite{bozzola_does_2022}, charge can leave imprints in key quantities at non-zero impact parameter near the scattering threshold. Does charge matter in other aspects of BBH dynamics, particularly in regards to the extremality of merger remnants and the energy and angular momentum radiated at moderate Lorentz factors? We begin to tackle these questions in this work.

We conduct new simulations of uncharged and like-charged BBHs near the non-linear scattering regime in full Einstein-Maxwell theory. Each BH has an initial boost of $\gamma = 1.520$ ($v/c = 0.753$) for comparison to previous uncharged studies~\cite{sperhake_cross_2009,smith_high-energy_2024}.
We vary the  initial charge-to-mass ratio \(\lambda \in \{0.0, 0.1, 0.4, 0.6\}\) and choose initial coordinate separation $d/M_{\rm ADM}\simeq 62$. We also vary the impact parameter $b$ for each $\lambda$.  The primary difference between the current study and Paper I is that we systematically vary the impact parameter for every $\lambda$ in our set, and do so at a smaller initial coordinate separation than in Paper I. A summary of our results follows. 

As we discovered in Paper I, the values of the impact parameter thresholds for scattering, $b_{\rm scat}/M_{\rm ADM}$, 
and immediate merger, $b^*/M_{\rm ADM}$,
both decrease for larger initial $\lambda$. 
However, when we normalize $b$ by $M_{\rm irr}$, we recover the universality we discovered in Paper I: $b_{\rm scat}/M_{\rm irr}$ and $b^*/M_{\rm irr}$, which determine the zoom-whirl regime in impact parameter space, become independent of $\lambda$. The same holds true if we estimate the impact parameter as $\overline{b}=J_{\rm ADM}/P$, where $J_{\rm ADM}$ is the spacetime ADM angular momentum, and $P$ the initial linear momentum of each BH.

In this work, we also introduce a new ``impact parameter-like" diagnostic based on the spacetime specific angular momentum, $a \equiv  J_{\rm ADM}/M_{\rm ADM}$, where $J_{\rm ADM}$ stands for the ADM angular momentum of the binary. Note that in geometrized units, $a$ has units of length. The motivation for $a$ comes from the definition of the impact parameter of a particle on a geodesic with energy $E$ and angular momentum $L$ in Schwarzschild geometry: $b=L/E$. We find that when we normalize the value of $a$ at the scattering threshold, $a_{\rm scat}$, by $M_{\rm irr}$, then $a_{\rm scat}/M_{\rm irr}$ is also universal. The same is true for the value of $a$ at the immediate merger threshold, $a^*/M_{\rm irr}$. 

When it comes to remnant properties, we find that increasing $\lambda$ increases the maximum possible $\Upsilon_f$, while reducing the maximum possible dimensionless spin of the remnant $j_f$. The most extremal merger in this work has a Kerr-Newman parameter of $\Upsilon_f = 0.97$, respecting cosmic censorship. 


Using our simulation data, we explore ways to parametrize the initial data such that we can predict the impact parameters that yield the maximal $\Upsilon_f$ and the maximal $j_f$ for a given $\lambda$ (at our fixed initial linear momenta). We observe that the maximal $\Upsilon_f$ for each $\lambda$ occurs at the same (to within the sampling accuracy in $b$) value of our new initial specific angular momentum parameter, $a$, normalized by $M_{\rm irr}$: $a/M_{\rm irr}=1.715 \pm 0.005$. We also observe that the location of the maximal $j_f$ for every initial $\lambda$ is consistent with $b/M_{\rm irr}=4.585 \pm 0.005$. The universality with $\lambda$ of these normalized impact parameter values, which are different from the threshold values for scattering and immediate merger, exemplify further that the areal radius of the BH, encoded in the irreducible mass, sets a fundamental length scale governing extreme BH encounters in horizon scale scattering events.

For each set of BBHs with the same value of $\lambda$, there is also a binary that radiates the most energy. We find that the value of this maximal radiated energy decreases with increasing $\lambda$. The maximum luminosity observed in our study (summing the electromagnetic and gravitational contributions) is $\frac{dE}{dt} = 0.023$ in geometrized units. Thus, our results respect the Dyson luminosity limit~\cite{dyson_ch_1963} (see also~\cite{cardoso_maximum_2018}). On the other hand, the maximum angular momentum radiated in our study exhibits very weak dependence on $\lambda$, and is $\approx 72\%$ of the spacetime total angular momentum.

In conclusion, we find that charge leaves imprints on key quantities near the scattering threshold for an initial Lorentz factor of $1.52$. Although this Lorentz factor is not in the ultrarelativistic regime, the head-on collisions in~\cite{bozzola_does_2022} showed that charge plays practically no role even at such low Lorentz factors. Far from the scattering threshold, the initial charge-to-mass ratio has little effect, but the charge-to-mass ratio of the binaries affects the values of $b_{\rm scat}/M_{\rm ADM}$, $b^*/M_{\rm ADM}$, the energy radiated, and the extremality of the merger remnants near but below the scattering threshold. It remains an open question as to whether charge will have an impact at larger initial Lorentz factors. We plan to explore this in future work.

The paper is structured as follows: in Sec.~\ref{sec:Methods} we describe the initial data, evolution methods, and diagnostics we adopt. We detail the results of our simulations in Sec. \ref{sec:Results}. We conclude in Sec. \ref{sec:Conclusions} with a summary and discussion. Throughout this paper we adopt geometrized units in which $G=c=(4\pi\epsilon_0)^{-1}=1$, where $G$ is the gravitational constant, $c$ the speed of light, and $\epsilon_0$ the vacuum permittivity.

\section{\label{sec:Methods}Methods}

We use the same methods for the generation of initial data, spacetime evolution, and diagnostics as in Paper I, where we refer the reader for more details. The infrastructure and many of the codes we adopt have been thoroughly tested and used in similar studies in the past~\cite{sperhake_cross_2009,sperhake_universality_2013}.

Our simulations are performed within the \texttt{Einstein Toolkit} infrastructure~\cite{brandt_einstein_2021} which uses {\tt Cactus}~\cite{Goodale2002a}. We adopt \texttt{TwoChargedPunctures}~\cite{bozzola_initial_2019} to generate our initial data. We use two codes from the \texttt{Canuda} suite~\cite{witek_canuda_2021}, \texttt{LeanBSSNMoL}~\cite{sperhake_binary_2007} and \texttt{ProcaEvolve}~\cite{zilhao_nonlinear_2015}, to evolve the spacetime and electromagnetic field, respectively. Adaptive mesh refinement is performed using \texttt{Carpet}~\cite{schnetter_evolutions_2004}. We locate apparent horizons with \texttt{AHFinderDirect}~\cite{thornburg_fast_2003} and calculate BH properties with \texttt{QuasiLocalMeasuresEM}~\cite{bozzola_initial_2019}, our version of \texttt{QuasiLocalMeasures}~\cite{dreyer_introduction_2003}. We use \texttt{kuibit}~\cite{bozzola_kuibit_2021} to analyze our simulation data. 

\subsection{\label{subsec:ID}Initial Data} 
Here we adopt a smaller initial coordinate separation than Paper I with $d/M_p = 94.85$ ($M_p=1.0$ is the sum of the target quasilocal gravitational masses of the BHs, see Paper I for more details). We do this for two reasons: i) we want to test the universality of the irreducible mass found in Paper I at a different initial separation, and ii) binary black holes with larger $\lambda$ require higher resolution, thereby making simulations more computationally expensive. To offset this increased cost, we set the initial separation in this work smaller than in Paper I. Here, we also expand the range of initial charge-to-mass ratios, investigating $\lambda = \{ 0.0, 0.1, 0.4, 0.6\}$ and impact parameters in the range $1.8 \leq b/M_{\rm ADM} \leq 3.3$. A detailed description of initial data parameters for our simulations is presented in Appendix \ref{sec:list_id}.
 
When setting up our initial data, we fix the magnitude of the initial linear momentum of each BH to $|P| = 0.57236$ and vary the BH charge in $Q^{1,2} = \{0.0, 0.05, 0.2, 0.3\}$. 
We set the target BH quasilocal mass to $M_p^{1,2} = 0.5$, the target charge-to-mass ratio to $\lambda^{1,2}=\lambda = \{0.0, 0.1, 0.4, 0.6\}$, 
and the target initial Lorentz factor to $\gamma^{1,2}=\gamma = 1.520$. The actual initial quasilocal BH masses $M^{1,2}$ --- calculated by \texttt{TwoChargedPuncturesEM} after initial data relaxation --- equal the target value of $M_p^{1,2} = 0.5$ to within $0.3 \%$ across our initial data. However, for a given $\lambda$, the initial $M^{1,2}$ are fixed to within $2$ parts in $10^5$ as $b$ is varied. The small variations in $M^{1,2}$ lead to small variations in $M_{\rm ADM}$, and the precise initial charge-to-mass ratio is $\overline{\lambda} \equiv Q^1/M^1=Q^2/M^2$. We list the precise values of these and other quanties in App.~\ref{sec:list_id}. Given that $\overline{\lambda}$ is very close to $\lambda$, we always refer to the BBHs by $\lambda$ outside of App.~\ref{sec:list_id},. 

As in Paper I, each simulation has three sets of nested refinement levels, two of which remain centered on the punctures and one on the origin. We have \(10\) or $11$ refinement levels, depending on the initial charge-to-mass ratio of the binary and the initial coordinate separation of the BHs. Our resolution varies with $\lambda$ to adjust for the change in apparent horizon size. The resolution is chosen such that the smallest coordinate radius of the BH apparent horizons is resolved by \(\sim 33\) grid-points after the initial data relaxes to the evolution gauge. Given this, our finest resolutions here are \(\{M_p/89, \, M_p/91, \, M_p/98, \, M_p/114\}\) for \(\lambda = \{0.0, \, 0.1, \, 0.4, \, 0.6\}\), respectively. Our outer grid boundary is placed at least $500 M_p$ from the origin for $\lambda = 0.0, 0.1$ and at least $850 M_p$ from the origin for $\lambda = 0.4, 0.6$.

\subsection{\label{subsec:diag} Diagnostics}
We employ the same diagnostic codes and tools as in Paper I, where we refer the reader for more details. 
We adopt \texttt{NPScalarsProca}~\cite{witek_black_2010, zilhao_nonlinear_2015} to extract the electromagnetic and gravitational radiation via the Weyl Newman-Penrose scalar $\Psi_4$ and Maxwell Newman-Penrose scalars $\phi_1$ and $\phi_2$.  

We use the Newman-Penrsose scalars to compute the energy and angular momentum radiated via gravitational waves (GWs) and electromagnetic waves (EMWs). We calculate the energy radiated via GWs and EMWs with Eqs.~19a \& 19b of \cite{bozzola_numerical-relativity_2021} up to and including $\ell = 6$ in the time domain. While integrating in the time domain, we find that removing the initial ``junk" radiation by cropping and windowing the $\Psi_4$ and $\phi_2$ data can cause artifacts in the luminosity. Instead, we implement the formulae without any cropping or windowing, and choose time bounds over which to integrate the GW and EMW luminosities. The junk radiation is identified as a burst in the luminosity near $(t - r) \approx 0$, so our lower bound is chosen to coincide with the local minimum immediately following the junk radiation peak. This minimum coincides across both the GW and EMW luminosities, and although it is not exactly zero, it is much, much smaller than the peak luminosity. Therefore, beginning our time integration at that location has negligible contribution to the total radiated energy. Our upper bound is selected to match when noise begins to dominate the $2,2$ mode of $\Psi_4$ and the GW and EMW luminosities have reached a steady value. Lastly, to mitigate artifacts generated by the time integration of the different $\ell,m$ modes of the spin-weighted spherical harmonic decomposition of $\Psi_4$, for each mode we subtract the integration constants~\cite{bishop_extraction_2016}. This forces the GW luminosity to zero at the upper boundary, removing any linear drifts in the total emitted energy with time. 

We tested our method by using our BBH with $\lambda=0.0$ and $b/M_p=  4.165$, which we estimate is closest to the $b/M=2.74$ case of ~\cite{sperhake_cross_2009} (here $M$ has the interpretation in that reference). Reference~\cite{sperhake_cross_2009} found this binary radiates 6.8\% of its total energy, and we find our BBH radiates $7.0\%$ of its total energy, in very good agreement with~\cite{sperhake_cross_2009}.

We calculate the angular momentum radiated via GWs with Eq.~19c of \cite{bozzola_numerical-relativity_2021} up to and including $\ell = 6$ in the time domain. The identification of the integration bounds for the angular momentum calculation requires slight modifications from the energy calculation integration bounds, the details of which we include in App.~\ref{sec:Rad_cutoffs}. We use the same method to remove the integration constants.   

The angular momentum radiated via EMWs is calculated with Eq.~17 of \cite{bozzola_numerical-relativity_2021}. We first recover the values of the $\phi_1$ and $\phi_2$ Newman-Penrose scalars from their spin-weighted spherical harmonic modes $\phi_1^{\ell,m}$ and $\phi_2^{\ell,m}$ across a spherical surface at the extraction radius, using up to and including $\ell = 6$. We then implement Eq.~17 of \cite{bozzola_numerical-relativity_2021} and numerically integrate over the surface to find the angular momentum radiated per unit time. Finally, we integrate in the time domain over the same bounds as the GW angular momentum calculation.  

The merger remnant properties are extracted with \texttt{QuasiLocalMeasuresEM} at  $ t/M_{\rm ADM}  \simeq 120$ after merger for each binary. The Kerr-Newman parameter of the merger remnant \(\Upsilon_f\) is computed  with Eq.~\eqref{eq:chif_def} from the charge \(Q_f\), gravitational mass \(M_f\), and spin angular momentum \(J_f\) of the remnant as calculated by \texttt{QuasiLocalMeasuresEM} via Eq.~A22 of \cite{bozzola_initial_2019}. Note that \(Q_f\) equals the sum of the initial BH charges to very high accuracy due to charge conservation. The maximum deviation between the sum of the initial charges and the charge of the merger remnant in our simulations is $0.002 \%$.

\section{\label{sec:Results}Results} 

\subsection{\label{subsubsec:thresholdSet2}Universality in Threshold Impact Parameters}
\begin{table}
\caption{\label{tab:sp_bvals}
Estimates of \(b^*\) and \(b_{\rm scat}\) for the different $\lambda$. The values are in agreement across $\lambda$ when normalized by \(M_{\rm irr}\). We also include an additional estimate of the impact parameter, $\overline{b} \equiv J_{\rm ADM}/|P|$, and the specific angular momentum, $a \equiv J_{\rm ADM}/M_{\rm ADM}$.
}
\begin{ruledtabular}
\begin{tabular}{cccc}
 $\lambda$ & 0.1 & 0.4 & 0.6\\
 \hline
 $b^*/M_{\rm ADM}$ & $3.24 \pm 0.03$ & $3.10 \pm 0.01$  & $2.90 \pm 0.01$\\
 $b^*/M_{\rm irr} $ & $4.95 \pm 0.04$ &$4.95 \pm 0.02$ & $4.97 \pm 0.02$\\
 $a^*/M_{\rm irr} $ & $1.85 \pm 0.02$ &$1.85 \pm 0.01$ & $1.84 \pm 0.01$\\
 $\overline{b}^*/M_{\rm ADM}$ & $3.23 \pm 0.03$ & $3.10 \pm 0.01$  & $2.90 \pm 0.01$\\
 $\overline{b}^*/M_{\rm ADM}$ & $3.23 \pm 0.03$ & $3.10 \pm 0.01$  & $2.90 \pm 0.01$\\
 $\overline{b}^*/M_{\rm irr} $ & $4.94 \pm 0.04$ &$4.95 \pm 0.02$ & $4.96 \pm 0.02$\\
 $b_{\rm scat}/M_{\rm ADM}$  & $3.29 \pm 0.02$ & $3.13 \pm 0.02$ & $2.96 \pm 0.03$\\
 $b_{\rm scat}/M_{\rm irr}$  &$5.02 \pm 0.04$ & $5.00 \pm 0.03$ & $5.07 \pm 0.04$\\
 $a_{\rm scat}/M_{\rm irr}$  &$1.88 \pm 0.01$ & $1.86 \pm 0.01$ & $1.87 \pm 0.02$\\
 $\overline{b}_{\rm scat}/M_{\rm ADM}$  & $3.28 \pm 0.02$ & $3.13 \pm 0.02$ & $2.96 \pm 0.03$\\
 $\overline{b}_{\rm scat}/M_{\rm irr}$  &$5.02 \pm 0.04$ & $5.00 \pm 0.03$ & $5.06 \pm 0.04$\\
\end{tabular}
\end{ruledtabular}
\end{table}

Our simulations recover and extend the universality with $M_{\rm irr}$ we found in Paper I: while the values of $b^*/M_{\rm ADM}$ and $b_{\rm scat}/M_{\rm ADM}$ decrease as $\lambda$ increases (showing dependence on charge), $b^*/M_{\rm irr}$ and $b_{\rm scat}/M_{\rm irr}$ are independent of $\lambda$ to within the determination error due to finite sampling in $b$. This consolidates the importance of $M_{\rm irr}$, which is proportional to the areal radius of the horizon, as a fundamental length scale that determines the outcome of horizon scale BH scattering events. The same result holds when we estimate the impact parameter as $\overline{b}=J_{\rm ADM}/|P|$. The values of $b^*$ and $b_{\rm scat}$, calculated with the approach described in Paper I and normalized by $M_{\rm ADM}$ and $M_{\rm irr}$, are reported Table~\ref{tab:sp_bvals}. 

We note that while the trends for the scattering and immediate merger thresholds agree between Paper I and here, the values of $b^*$ and $b_{\rm scat}$ for a given $\lambda$ do not exactly agree, indicating an additional dependence on initial BH coordinate separation. This is likely due to the fact that formally $b$ is defined at very large separation, and one needs to account for propagation effects to the smaller initial separations we adopt. In our setup (see Fig.~1 in Paper I), the relation between the impact parameter, initial coordinate separation, and momentum is given by
\begin{eqnarray} 
P_y = \left|P\right| \frac{b}{d} \, ; \,\,\,\,\,\,\, P_x =  \sqrt{P^2 - P_y^2}\, 
\label{eq:PxPy}.
\end{eqnarray}
Based on Eq.~\eqref{eq:PxPy}, the larger the initial separation, the closer the value of $b$ (defined at that finite separation) is to the true value, formally defined at $b/d\ll 1$, because $P_y/\left|P\right|\ll 1$ when $b\ll d$. Thus, at finite separation one expects  corrections of order $b/d$. In Paper I,  we had $b/d = O(1\%)$. However, here, $b/d = O(5\%)$. Coincidentally, the values of $b^*(\lambda)$ (and $b_{\rm scat}(\lambda)$) in Paper I and here differ by $O(3\%$).

An additional ``impact parameter-like" diagnostic is listed in Table~\ref{tab:sp_bvals}: the specific angular momentum $a \equiv J_{\rm ADM}/M_{\rm ADM}$. This diagnostic is useful in later scaling of remnant extremality curves. Note that the values of $a^*/M_{\rm irr}$ and $a_{scat}/M_{\rm irr}$ are also independent of $\lambda$.

\subsection{Properties of Merger Remnants}

\begin{figure}
\includegraphics[width=0.47\textwidth]{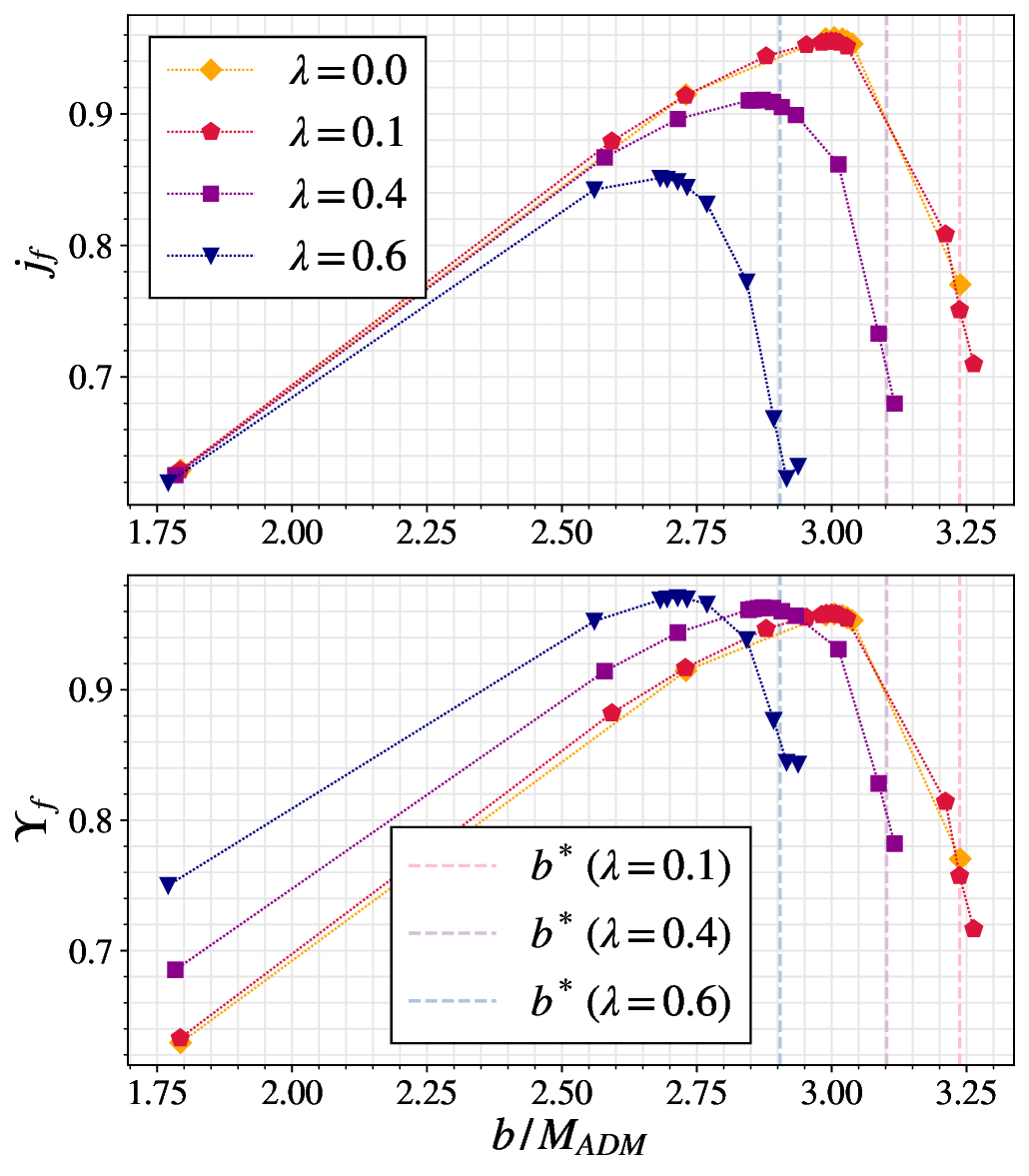}
\caption{\label{fig:extr} Top panel: remnant dimensionless spin $j_f$ vs $b/M_{\rm ADM}$. Bottom panel: remnant Kerr-Newman extremality parameter $\Upsilon_f$ vs $b/M_{\rm ADM}$.  The line colors and marker styles differentiate the different $\lambda$, as described in the legend. The vertical dashed lines in both panels represent the immediate merger threshold $b^*$ for each $\lambda$, as indicated in the legend of the bottom panel. }
\end{figure}

In Fig.~\ref{fig:extr} we show the dimensionless spin $j_f$ (top) and Kerr-Newman  parameter $\Upsilon_f$ (bottom) of the merger remnants as functions of $b/M_{\rm ADM}$ and $\lambda$. 
We notice that for $\lambda\geq 0.4$, charge has a substantial effect on $j_f$ at the same value of $b$ for $b/M_{\rm ADM} \gtrsim 2.7$. For example, $j_f$ for $\lambda = 0.6$ can be $\sim 35\%$ smaller than for $\lambda = 0.0$ at the same impact parameter (Fig.~\ref{fig:extr}, top panel). 

The strong dependence of $j_f$ on the initial charge-to-mass ratio in this regime indicates that charge has a marked effect on the BBH dynamics as $b/M_{\rm ADM}$ approaches $b^*$. Rather than charge simply ``coming along for the ride" and only contributing a greater $\lambda_f$ to the extremality of the merger remnant, it is modifying the energy and angular momentum radiated. This modification is seen in the variation of $j_f$ (Fig. \ref{fig:extr}, top panel). In other words, charge leaves a significant imprint as $b$ approaches $b^*$ for the moderate Lorentz factor $\gamma = 1.520$. This is in contrast to head-on collisions at moderate Lorentz factors such as the one adopted in our work, where charge was shown to be unimportant~\cite{bozzola_does_2022}.

While $\lambda$ has a marked effect on $j_f$ once $b$ reaches $ \sim 90 \% \, b^*$, it has a nearly negligible effect at small $b/M_{\rm ADM}$, with \(j_f\) differing by less than \(1.5\%\) across $\lambda$ for our smaller $b$ simulations. Thus, our simulations generalize the result of~\cite{bozzola_does_2022} that charge does not matter in BH encounters at moderate $\gamma$ to non-zero impact parameter, but for impact parameters a few times smaller than $b^*$.

Moreover, the top panel in Fig.~\ref{fig:extr} shows that the maximum of the $j_f$ vs $b/M_{\rm ADM}$ curves, and the value of $b/M_{\rm ADM}$ where this maximum occurs both decrease as $\lambda$ increases. This is consistent with our results from Paper I, where we observed that $b^*/M_{\rm ADM}$ and $b_{scat}/M_{\rm ADM}$ decrease as $\lambda$ increases. 
\begin{table}
\caption{\label{tab:extr_vals}
Impact parameter $b$, Kerr-Newman extremality parameter $\Upsilon_f$, dimensionless spin $j_f$, and charge-to-mass ratio $\lambda_f$ of the most extremal merger remnant for each $\lambda$.}
\begin{ruledtabular}
\begin{tabular}{ccccc}
 $\lambda$ & $b/M_{\rm ADM}$  & $\Upsilon_f$ & $j_f$ & $\lambda_f$ \\
 \hline
 0.0 & 3.00 & 0.96 & 0.96 & 0.00 \\
 0.1 & 3.00 & 0.96 & 0.96 & 0.08 \\
 0.4 & 2.87 & 0.96 & 0.91 & 0.31 \\
 0.6 & 2.71 & 0.97 & 0.85 & 0.47 \\
\end{tabular}
\end{ruledtabular}
\end{table} 

The bottom panel of Fig.~\ref{fig:extr} shows that the most extremal remnant for each $\lambda$ always occurs at an impact parameter below $b^*(\lambda)$. The properties of the most extremal remnants for each $\lambda$ in our set are listed in Table~\ref{tab:extr_vals}, which shows that we can always find remnants with $\Upsilon_f \geq 0.96$, and the most extreme remnant in this study has \(\Upsilon_f = 0.97\) and  \(j_f = 0.85\). Thus, we find that cosmic censorship is respected. 

In Fig.~\ref{fig:extr} we also notice that for a given $\lambda$ the maximum of the $j_f$--$b/M_{\rm ADM}$ and maximum of the $\Upsilon_f$--$b/M_{\rm ADM}$ curves occur at different values of $b$. This feature is a consequence of the definition of $\Upsilon_f$, and we explain it in App.~\ref{app:peaks_exp}.

\begin{figure}
\includegraphics[width=0.47\textwidth]{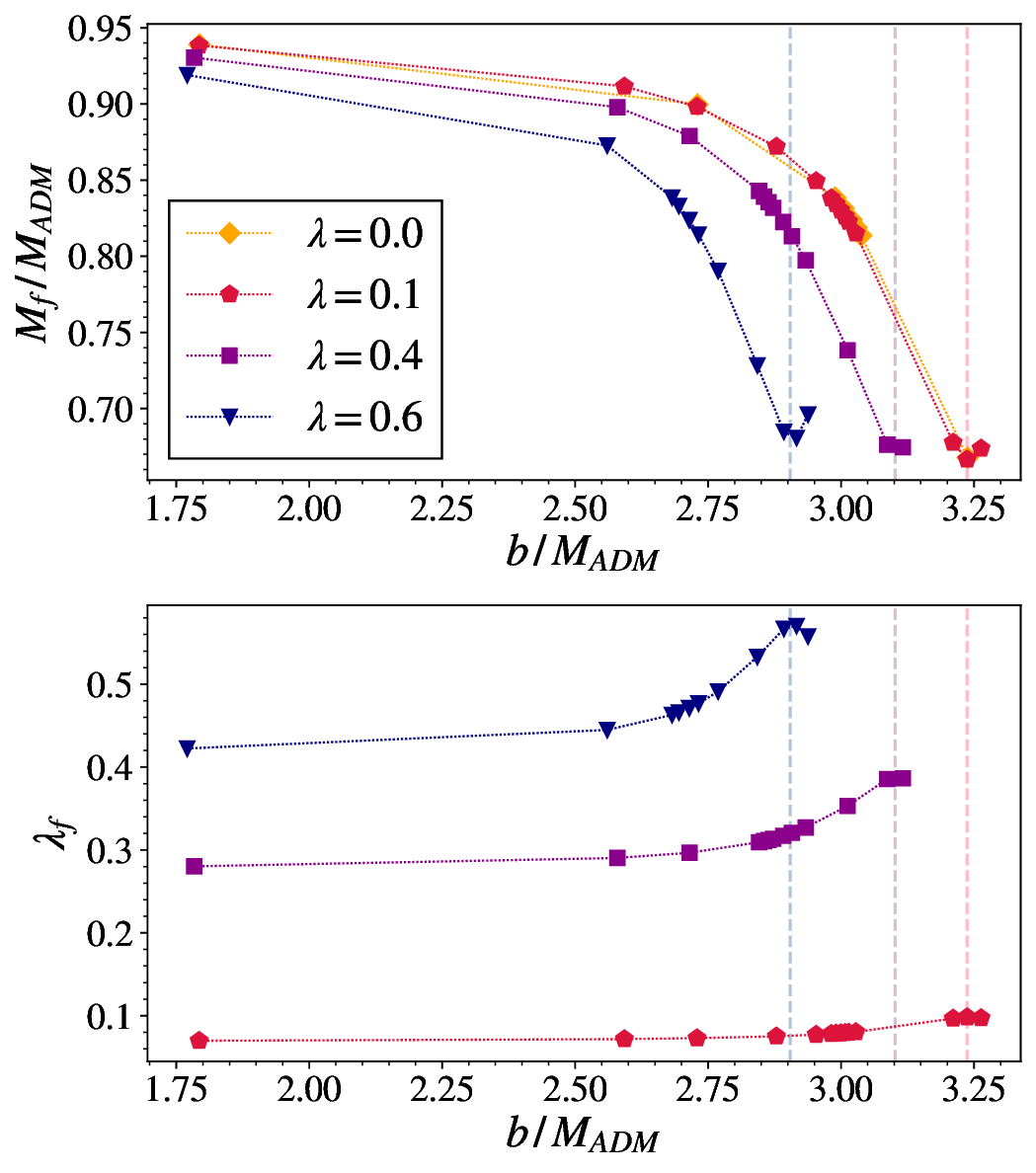}
\caption{\label{fig:lam_M} Top panel: The remnant mass $M_f$ normalized by $M_{\rm ADM}$ vs $b/M_{\rm ADM}$. Bottom panel: the remnant $\lambda_f$ vs $b/M_{\rm ADM}$. The legend in the top panel applies to both plots, and the dashed lines indicate the value of $b^*$ for a given $\lambda$ (denoted by the color-coding shown in the legend).}
\end{figure}

In the top panel of Fig.~\ref{fig:lam_M}, we show the remnant mass $M_f$ as we vary $\lambda$ and $b/M_{\rm ADM}$. For any given $\lambda$, we notice that $M_f$ monotonically decreases with increasing $b/M_{\rm ADM}$ when $b<b^*$ . This is consistent with the rise in radiated energy as $b/M_{\rm ADM}$ increases for $b<b^*$ that we discuss in Sec.~\ref{subsubsec:rad} below. Furthermore, for the range of $b/M_{\rm ADM}$ in our $\lambda=0.6$ case, the ratio $M_f / M_{\rm ADM}$ decreases as $\lambda$ increases. 

As $b$ decreases well below $b^*$, the effect of $\lambda$ on $M_f$ is consistent with the $\Upsilon_f$ results, i.e., charge has a very small effect. In particular, at our smallest impact parameter $b/M_{\rm ADM} = 1.8$, the value of $M_f / M_{\rm ADM}$ for $\lambda = 0.6$ and $\lambda = 0.0$ differ by only $2.1 \%$. Similarly to $\Upsilon_f$, $M_f / M_{\rm ADM}$ depends strongly on $\lambda$ when $b/M_{\rm ADM} > 2.75$, suggesting that the initial charge-to-mass ratio influences the dynamics of the system in this regime.

The bottom panel of Fig.~\ref{fig:lam_M} plots the remnant charge-to-mass ratio $\lambda_f$ as a function of $\lambda$ and $b/M_{\rm ADM}$. Binaries with larger initial $\lambda$ have larger $\lambda_f$, and we always observe that $\lambda_f\leq \lambda$. While this may appear contradictory with the fact that mass-energy is radiated away and charge is conserved, we note that the mass that goes into the definition of $\lambda$ does not account for the Lorentz factor; the ADM mass of the spacetime is larger than the sum of the gravitational masses (not accounting for linear momentum) of the initial BHs. In other words, the $\lambda$ for the initial BHs would be the true charge-to-mass ratio if the initial BHs were at rest at infinity. By contrast, $\lambda_f$ is the charge-to-mass ratio of the remnant BH, because the remnant has zero linear momentum. 

We note that all BBH remnants with a given $\lambda$ have the same total charge $Q_f$ regardless of $b$ because of charge conservation. Therefore, when plotting $\lambda_f = Q_f / M_f$ versus $b/M_{\rm ADM}$, only $M_f$ changes with $b$. 

\begin{figure}
\includegraphics[width=0.4\textwidth]{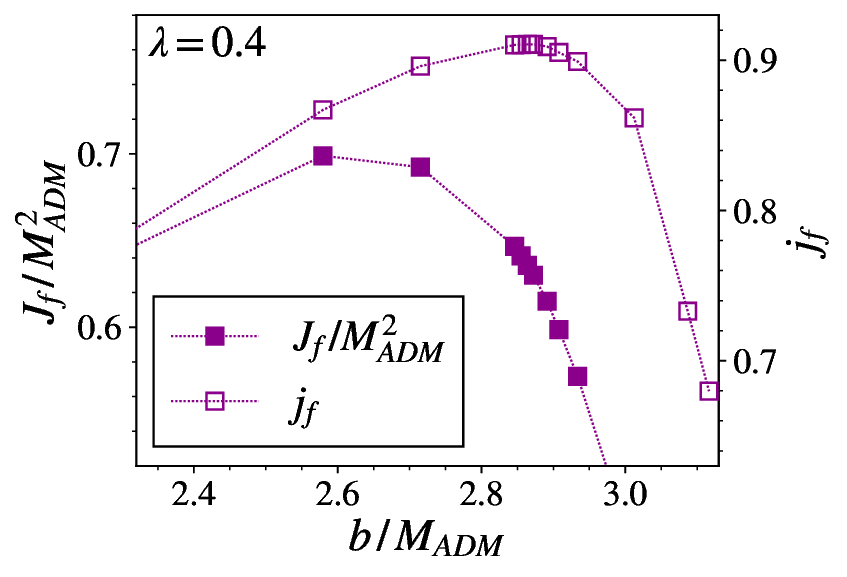}
\caption{\label{fig:J_v_j} Merger remnant dimensionless spin $j_f$ (empty markers, right axis) and angular momentum $J_f$ (solid markers, left axis) with respect to $b/M_{\rm ADM}$ for $\lambda = 0.4$. The plot is zoomed in around the peaks of both curves. $J_f$ is normalized by the ADM mass $M_{\rm ADM}$, which is constant for all practical purposes.}
\end{figure}

Our simulations help illuminate the interplay between charge, mass, and angular momentum discussed in the introduction. The large portion of energy radiated via gravitational and electromagnetic waves (see Sec. \ref{subsubsec:rad}) causes $M_f$ to decrease with increasing $b$ for $b<b^*$ (Fig.~\ref{fig:lam_M}, top panel). This variation in $M_f$ affects the dimensionless spin: Fig.~\ref{fig:J_v_j} 
plots the spin angular momentum $J_f$ and dimensionless spin $j_f = J_f/M_f^2$ of the $\lambda = 0.4$ merger remnants as functions of $b$, zoomed in around the peak in both curves.\footnote{The $\lambda = 0.4$ case is representative for all initial non-zero $\lambda$. We do not include all cases, because the spread in peak locations across $\lambda$ would make the plot difficult to parse.} The spin angular momentum $J_f$ is normalized by $M_{\rm ADM}^2$, which varies by at most $2$ parts in $10^6$ as $b$ is changed (for $\lambda = 0.4$). Thus, for all practical purposes in this plot, $M_{\rm ADM}$ is a constant and is just used to normalize $J_f$.  The plot demonstrates that as $b$ increases from small values, $J_f$ initially increases with $b$ and so does $j_f$. As $b$ increases past $\simeq 0.8b^*$, $J_f$ drops, but $j_f$ increases because $M_f$ drops faster than $J_f$. This continues until $b\simeq 0.9b^*$, after which the drop in $J_f$ is faster than the drop in $M_f$, and $j_f$ begins to decrease. A similar trend exists in the other $\lambda$ cases we study.

Figure~\ref{fig:J_v_j} also shows that  for each $\lambda$ explored here, the maximum $J_f$ and maximum $j_f$ 
occur at different values of $b/M_{\rm ADM}$. We explain this behavior in App.~\ref{app:peaks_exp}.

\begin{figure}
\includegraphics[width=0.45\textwidth]{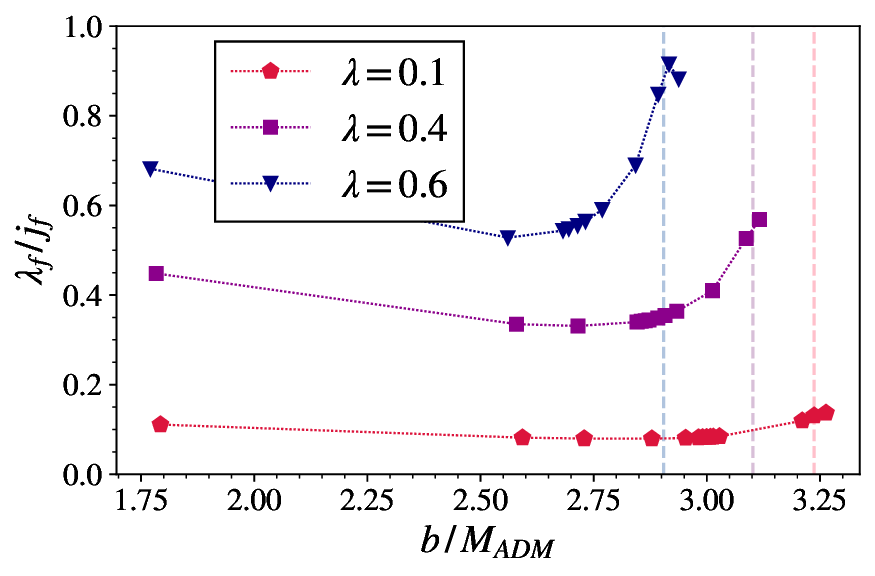}
\caption{\label{fig:lam_j_prop} The remnant charge-to-mass ratio $\lambda_f$, normalized by the corresponding $j_f$, vs $b/M_{\rm ADM}$ for different values of $\lambda$. The ratio $\lambda_f/j_f$ never reaches 1.0, indicating that $j_f$ always has the larger contribution to $\Upsilon_f$. 
The dashed vertical lines indicate the values of $b^*$, with $\lambda$ denoted by color.}
\end{figure}

To understand the relative contribution of angular momentum and charge to the extremality of our remnants, we show in Fig.~\ref{fig:lam_j_prop} the ratio of $\lambda_f/j_f$ as a function of $\lambda$ and $b/M_{\rm ADM}$. 
As expected, $\lambda_f /j_f$ increases with $\lambda$, with $\lambda_f/j_f < 1$ for all $\lambda$ and $b$ investigated here. This  demonstrates that $j_f$ has a larger contribution to $\Upsilon_f$ than $\lambda_f$ for $\lambda \leq 0.6$.  However, one of the $\lambda = 0.6$ remnants has $\lambda_f/j_f > 0.9$, indicating that binaries with $\lambda > 0.6$ may be able to reach $\lambda_f / j_f \geq 1.0$. 

\subsection{\label{subsubsec:ID_scalings}Merger Remnants: Universality with $M_{\rm irr}$}

In this section we seek universal, i.e., $\lambda$-independent, effective impact parameters that can align the peaks in the $j_f$ vs effective impact parameter curves, and align the peaks in the $\Upsilon_f$--effective impact parameter curves. We define two different effective impact parameters, which both use the sum of the initial irreducible masses $M_{\rm irr}$, providing further evidence for the potential ubiquity of the importance of the irreducible mass (or the horizon areal radius) in setting the fundamental scale in horizon scale BH interactions. 

\begin{figure}
\includegraphics[width=0.41\textwidth]{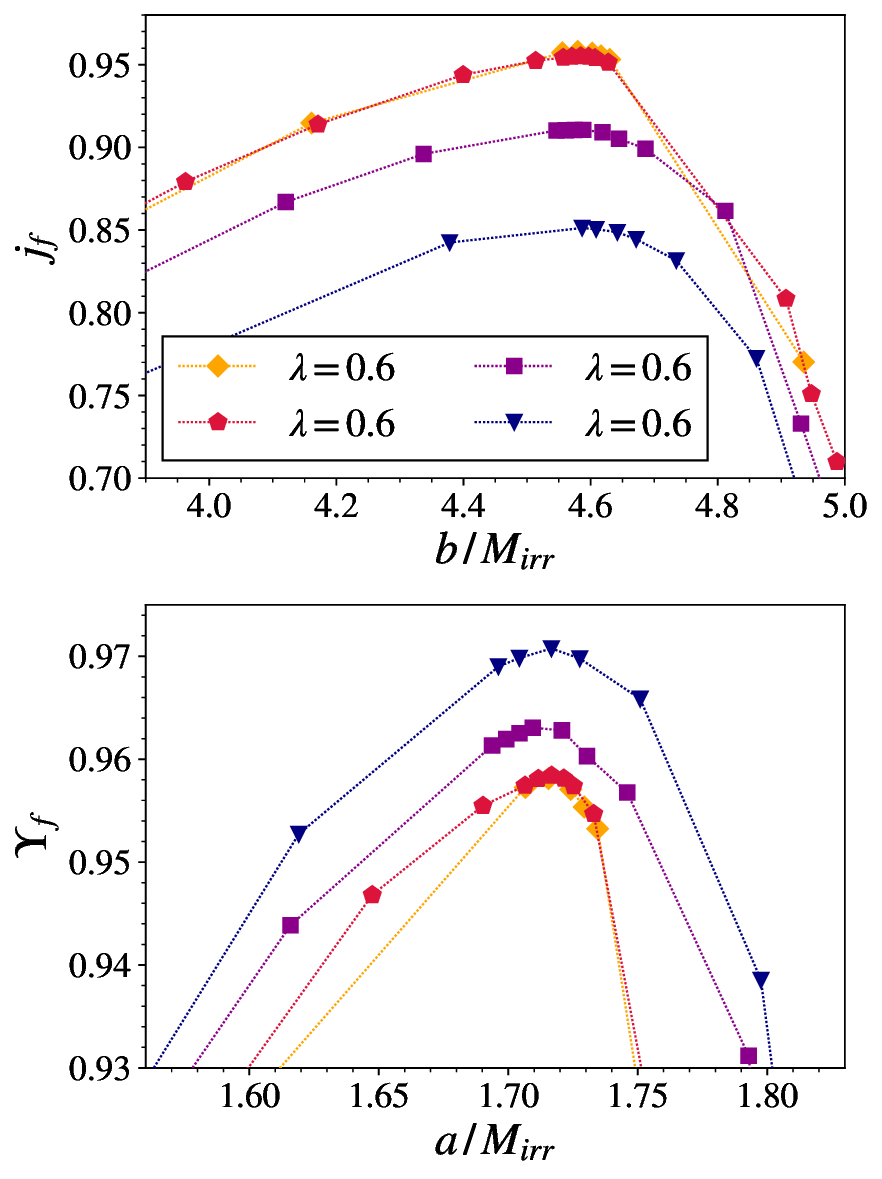}
\caption{\label{fig:chi_alpha} Top panel: Remnant dimensionless spin $j_f$ vs $b/M_{\rm irr}$. Bottom panel: Kerr-Newman parameter $\Upsilon_f$ vs $a/M_{\rm irr}$. The different charge-to-mass ratios $\lambda$ are shown with different line style and color as indicated in the legend of the top panel. }
\end{figure}
\begin{table}
\caption{\label{tab:b_max}
Values of 
$b/M_{\rm irr}$ and $b/M_{\rm ADM}$ for BBHs with the greatest $j_f$ and $\Upsilon_f$ in each $\lambda$. The error is the distance $ \Delta b/M_{\rm irr}$ or $ \Delta b/M_{\rm ADM}$ to the next highest values of $j_f$ or $\Upsilon_f$. We also include an additional estimate of the impact parameter, $\overline{b} \equiv J_{\rm ADM}/|P|$.}
\begin{ruledtabular}
\begin{tabular}{ccccc}
 $\lambda$ & $0.0$ & $0.1$ & $0.4$ & $0.6$ \\ 
 \hline 
 $b/M_{\rm irr}|_{j_f^{max}}$ & $4.58^{+0.02}_{-0.02}$ & $4.58^{+0.01}_{-0.01}$ & $4.59^{+0.03}_{-0.01}$ & $4.59^{+0.02}_{-0.2}$ \\
 $b/M_{\rm irr}|_{\Upsilon_f^{max}}$ & $4.58^{+0.02}_{-0.02}$ & $4.58^{+0.01}_{-0.01}$ & $4.59^{+0.03}_{-0.01}$ & $4.64^{+0.03}_{-0.03}$ \\
 $\overline{b}/M_{\rm irr}|_{j_f^{max}}$ & $4.57^{+0.02}_{-0.02}$ & $4.58^{+0.01}_{-0.01}$ & $4.58^{+0.03}_{-0.01}$ & $4.58^{+0.02}_{-0.2}$ \\
 $\overline{b}/M_{\rm irr}|_{\Upsilon_f^{max}}$ & $4.57^{+0.02}_{-0.02}$ & $4.58^{+0.01}_{-0.01}$ & $4.58^{+0.03}_{-0.01}$ & $4.63^{+0.03}_{-0.03}$ \\
 $b/M_{\rm ADM}|_{j_f^{max}}$ & $3.00^{+0.02}_{-0.02}$ & $3.00^{+0.01}_{-0.01}$ & $2.87^{+0.02}_{-0.01}$ & $2.68^{+0.01}_{-0.1}$ \\
 $b/M_{\rm ADM}|_{\Upsilon_f^{max}}$ & $3.00^{+0.02}_{-0.02}$ & $3.00^{+0.01}_{-0.01}$ & $2.87^{+0.02}_{-0.01}$ & $2.71^{+0.02}_{-0.02}$ \\
 $\overline{b}/M_{\rm ADM}|_{j_f^{max}}$ & $3.00^{+0.02}_{-0.02}$ & $2.99^{+0.01}_{-0.01}$ & $2.87^{+0.02}_{-0.01}$ & $2.68^{+0.01}_{-0.1}$ \\
 $\overline{b}/M_{\rm ADM}|_{\Upsilon_f^{max}}$ & $3.00^{+0.02}_{-0.02}$ & $2.99^{+0.01}_{-0.01}$ & $2.87^{+0.02}_{-0.01}$ & $2.71^{+0.02}_{-0.02}$ \\
 \end{tabular}
\end{ruledtabular}
\end{table} 

Figure~\ref{fig:chi_alpha} demonstrates the $\lambda$-independence of the maximum remnant dimensionless spin $j^{max}_f$ and Kerr-Newman parameter $\Upsilon_f$ as a function of impact parameter quantities normalized by $M_{\rm irr}$. The maximum remnant $j^{max}_f$ becomes independent of $\lambda$ when plotted as a function of $b/M_{\rm irr}$. In this case, all our simulations have maximum $j^{max}_f$ for $4.58 < b/M_{\rm irr} < 4.59$, as reported in Table~\ref{tab:b_max}, where we also list the error due to finite sampling over $b$.

While we find a high-level of universality between the dimensionless spin $j^{max}_f$ and $b/M_{\rm irr}$, the universality with $b/M_{\rm irr}$ of the $\Upsilon_f$ peaks is at best at the $0.4\%$ level. The universality can be improved by using an effective impact parameter $a\equiv J_{\rm ADM} / M_{\rm ADM}$ normalized by $M_{\rm irr}$. This is reported in the bottom panel of Fig.~\ref{fig:chi_alpha}, which shows $\Upsilon_f$ vs $a/M_{\rm irr}$ as a function of $\lambda$. The peaks of the function $\Upsilon_f(a/M_{\rm irr})$ are all in the range $ 1.71 \leq a/ M_{\rm irr} \leq 1.72$ (see Table~\ref{tab:a_max}). 

\begin{table}
\caption{\label{tab:a_max}
Values of
$a/M_{\rm irr}$ and $a/M_{\rm ADM}$ for BBHs with the maximum $j_f$ and $\Upsilon_f$ in each $\lambda$. The error is the distance $ \Delta a/M_{\rm irr}$ or $\Delta a/M_{\rm ADM}$ to the next highest values of $j_f$ or $\Upsilon_f$.}
\begin{ruledtabular}
\begin{tabular}{ccccc}
 $\lambda$ & $0.0$ & $0.1$ & $0.4$ & $0.6$ \\ 
 \hline
 $a/M_{\rm irr}|_{j_f^{max}}$ & $1.716^{+0.009}_{-0.009}$ & $1.717^{+0.005}_{-0.005}$ & $1.709^{+0.01}_{-0.005}$ & $1.696^{+0.008}_{-0.08}$ \\
 $a/M_{\rm irr}|_{\Upsilon_f^{max}}$ & $1.716^{+0.009}_{-0.009}$ & $1.717^{+0.005}_{-0.005}$ & $1.709^{+0.01}_{-0.005}$ & $1.72^{+0.01}_{-0.01}$ \\
 $a/M_{\rm ADM}|_{j_f^{max}}$ & $1.126^{+0.006}_{-0.006}$ & $1.123^{+0.003}_{-0.003}$ & $1.070^{+0.007}_{-0.003}$ & $0.992^{+0.005}_{-0.05}$ \\
 $a/M_{\rm ADM}|_{\Upsilon_f^{max}}$ & $1.126^{+0.006}_{-0.006}$ & $1.123^{+0.003}_{-0.003}$ & $1.070^{+0.007}_{-0.003}$ & $1.004^{+0.006}_{-0.007}$ \\
 \end{tabular}
\end{ruledtabular}
\end{table} 

\begin{figure}
    \includegraphics[width=0.48\textwidth]{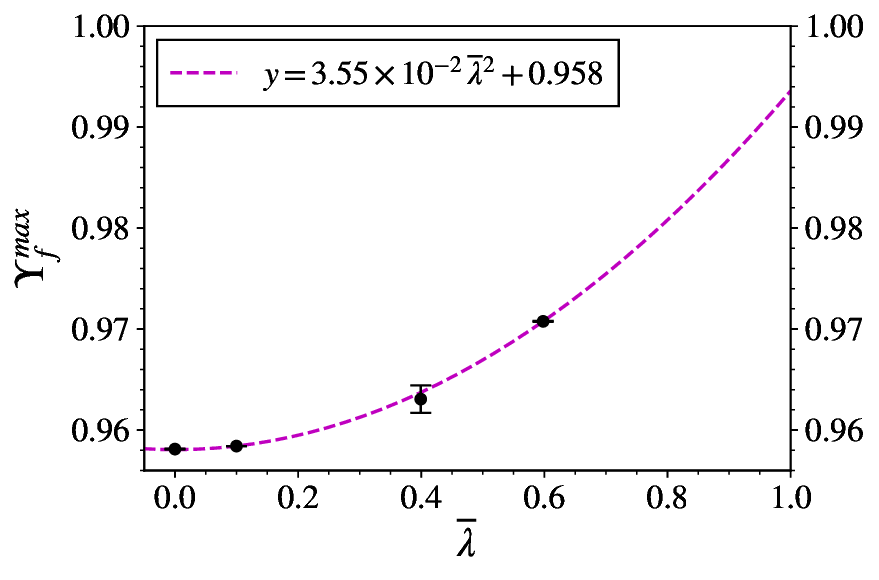}
    \caption{\label{fig:proj_chi}  The maximum $\Upsilon_f$ for each $\lambda$ as a function of $\overline{\lambda}$ (black points). A second-order polynomial best fit (magenta, dashed) is shown. Here, $\overline{\lambda}$ is the precise initial charge-to-mass ratio calculated via the isolated horizon formalism. The error bars are magnified by a factor of $10$ for visibility. A description of their calculation is included in App.~\ref{sec:Chi_err}.}
\end{figure}

The bottom panel of Fig.~\ref{fig:chi_alpha} also demonstrates that the maximum value of $\Upsilon_f$ for a given $\lambda$, which we designate with $\Upsilon_f^{\rm max}$ in the plot, increases as $\lambda$ increases. In Fig.~\ref{fig:proj_chi} we use a second-order polynomial in $\overline{\lambda}$ to fit $\Upsilon_f^{\rm max}$. We adopt an even order polynomial because our BHs have like charges, so the results should be invariant as $\overline{\lambda} \rightarrow - \overline{\lambda}$. 
Note that $\overline{\lambda}$ is the charge-to-mass ratio calculated with our BH diagnostics after initial data relaxation. Using the best-fit polynomial, we can extrapolate $\Upsilon_f^{\rm max}$ to $\overline{\lambda} = 1.0$ and find that $\Upsilon_f^{\rm max}=0.994$. Therefore, our extrapolation predicts that a naked singularity would not form for $\overline{\lambda} = 1$. We plan to test this tentative prediction in a future work.

The aforementioned fit was performed using the \texttt{curve\_fit} function from \texttt{scipy}~\cite{virtanen_scipy_2020}, which takes as input the error bars $\delta \Upsilon_f^{max}\left(\overline{\lambda}\right)$ shown in Fig.~\ref{fig:proj_chi}. A description of the calculation of $\delta \Upsilon_f^{max}\left(\overline{\lambda}\right)$ is included in App.~\ref{sec:Chi_err}. 

\subsection{\label{subsubsec:rad}Radiated Quantities}

\begin{figure}[h!]
\includegraphics[width=0.46\textwidth]{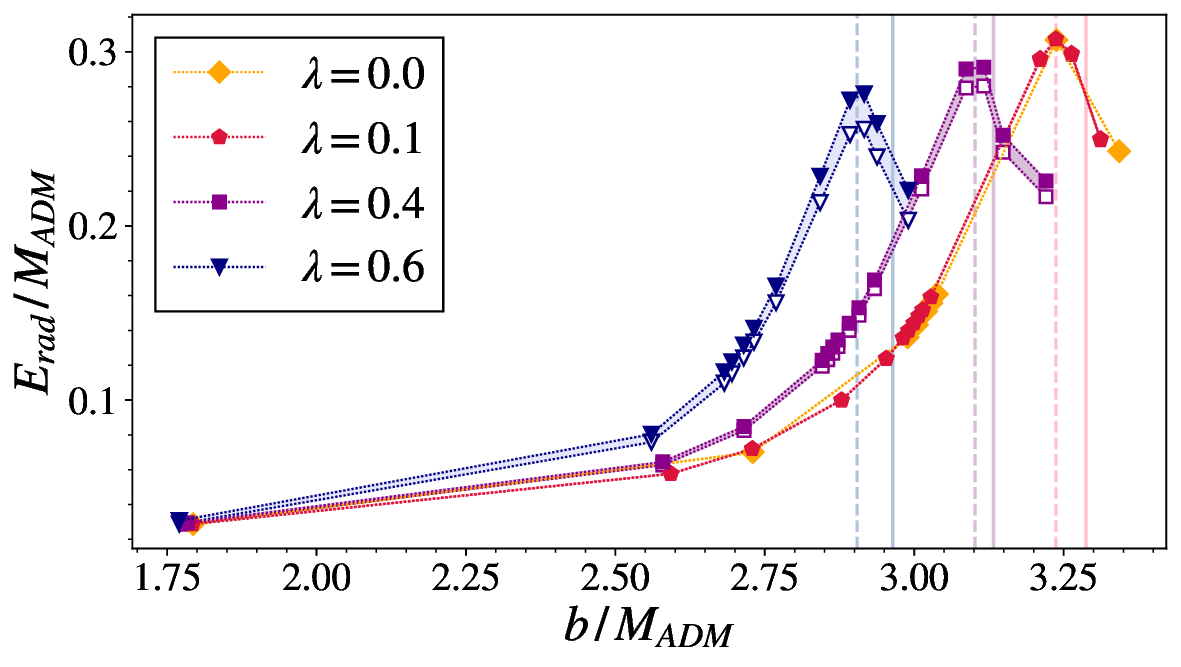}
\caption{\label{fig:E_rad} 
Fraction of the ADM mass radiated as a function of \(\lambda\) and \(b/M_{\rm ADM}\). The solid markers plot the sum of the energy radiated via GWs and EMWs. The empty markers plot the energy radiated via GWs. The vertical lines correspond to $b^*$ (dashed) and $b_{scat}$ (solid) for each $\lambda$, indicated by color.}
\end{figure}

In Fig.~\ref{fig:E_rad} we plot the energy radiated via gravitational waves (GWs) and electromagnetic waves (EMWs), normalized to $M_{\rm ADM}$, as a function of $\lambda$ and $b/M_{\rm ADM}$. 
The maximum total energy radiated  is \(E_{rad}/ M_{\rm ADM} = 0.31\), reached for \(\lambda = 0.0\) and $\lambda = 0.1$ at $b/M_{\rm ADM} = 3.24$. We find that \(E_{rad}\) is almost insensitive to $\lambda$ at low $b/M_{\rm ADM}$, but as $b/M_{\rm ADM}$ increases, \(E_{rad}\) displays a dependence on $\lambda$ at a given $b/M_{\rm ADM}$. We observe in Fig.~\ref{fig:E_rad} that both the maximum of the $E_{rad}/M_{\rm ADM}$--$b/M_{\rm ADM}$ curve for a given $\lambda$, and the value of $b/M_{\rm ADM}$ at which the maximum occurs, decrease as $\lambda$ increases. This gives further indication that as $b/M_{\rm ADM}$ approaches $b^*$ and $b_{\rm scat}$, charge leaves a measurable imprint.

As expected, greater initial charge-to-mass ratios radiate more energy via EMWs. (This can be seen by the width of the shading in Fig.~\ref{fig:E_rad}.) At the smallest impact parameter evaluated, $b/M_{\rm ADM} = 1.8$, the energy radiated via EMWs scales like $\lambda^2$, as expected. 
The value $b/M_{\rm ADM} = 1.8$ provides the best apples-to-apples comparison in our dataset, because we have data for $b/M_{\rm ADM} = 1.8$ for all $\lambda$ in our set, and because $b$ is not in the regime where charge can non-linearly affect the dynamics, thereby modifying $E_{rad}^{EMW}$. 

\begin{figure}
\includegraphics[width=0.46\textwidth]{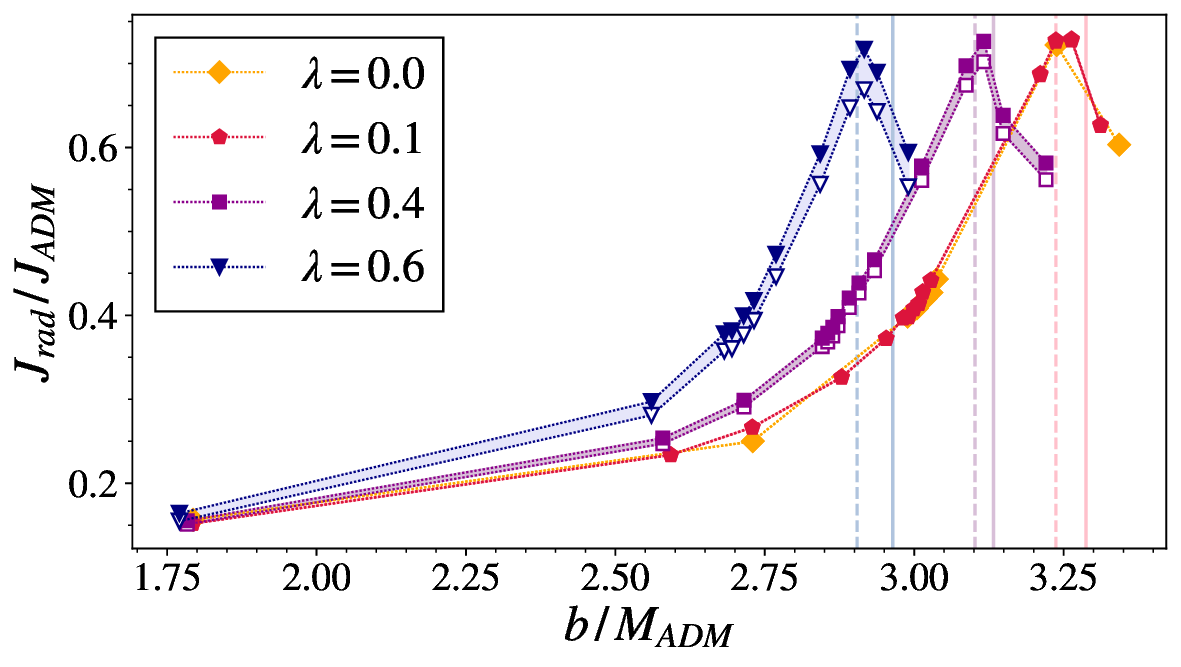}
\caption{\label{fig:J_rad} Fraction of ADM angular momentum radiated with respect to \(\lambda\) and \(b/M_{\rm ADM}\). The marker and linestyles have the same meanings as in Fig.~\ref{fig:E_rad}.}
\end{figure}

\begin{figure*}
\includegraphics[width=0.95\textwidth]{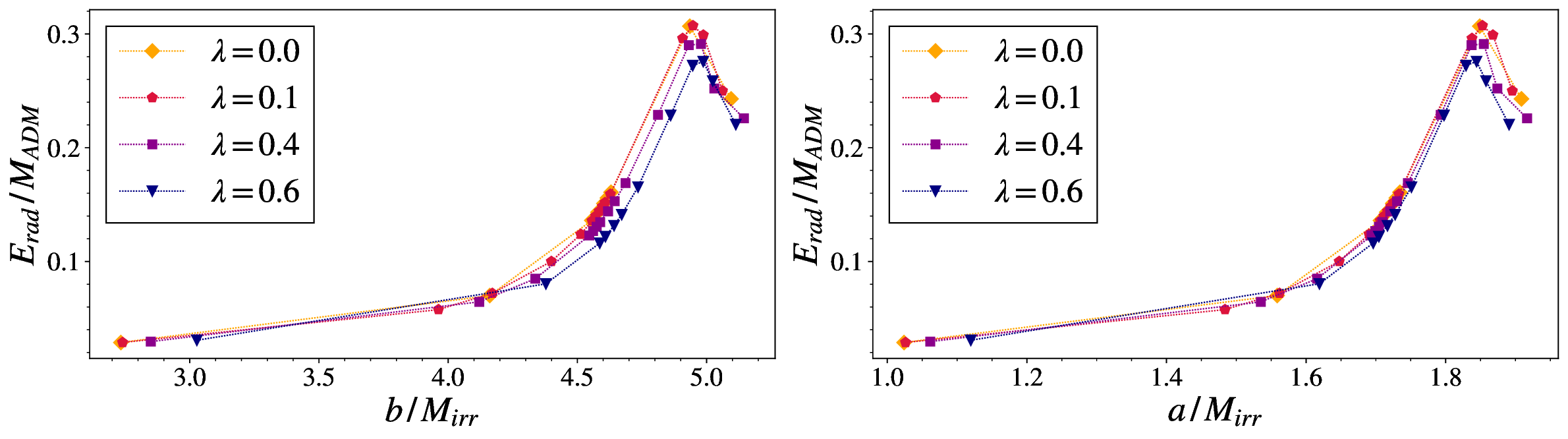}
\caption{\label{fig:Erad_combo} Portion of  ADM mass $M_{\rm ADM}$ radiated by each binary via GWs and EMWs. The left panel plots these values with respect to $b/M_{\rm irr}$, the right with respect to $a/M_{\rm irr}$. The curves represent different $\lambda$, as described by the legend.}
\end{figure*}

\begin{figure*}
\includegraphics[width=0.95\textwidth]{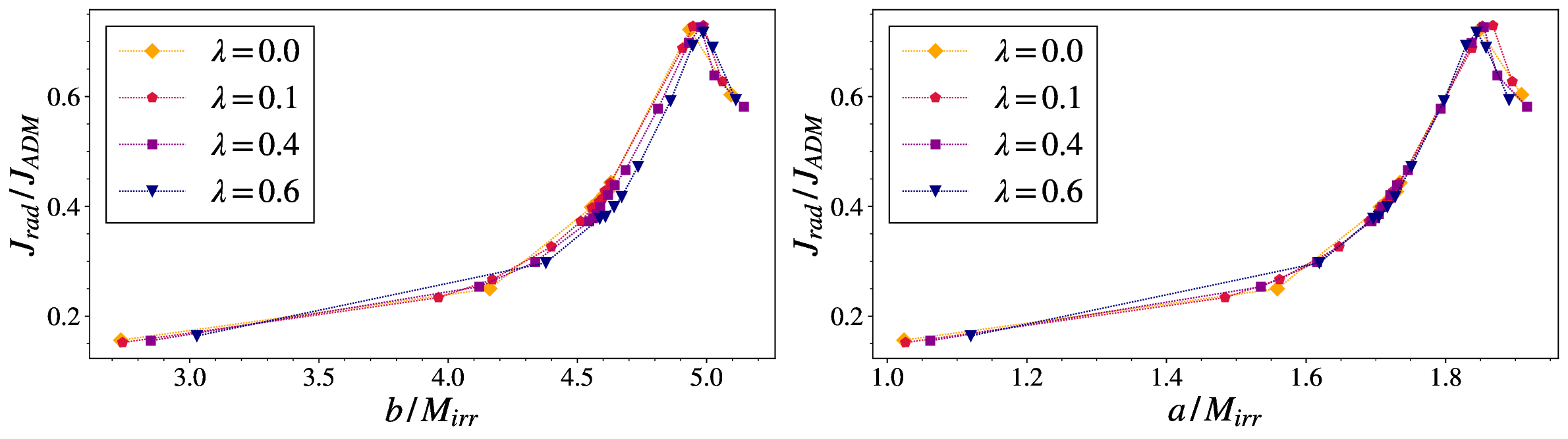}
\caption{\label{fig:Jrad_combo} Portion of  ADM angular momentum $J_{\rm ADM}$ radiated by each binary via GWs and EMWs. The left panel plots these values with respect to $b/M_{\rm irr}$, the right with respect to $a/M_{\rm irr}$. The curves represent different $\lambda$, as described by the legend. }
\end{figure*}

The vertical lines in Fig.~\ref{fig:E_rad} designate the values of $b^*/M_{\rm ADM}$ (dashed lines) and $b_{scat}/M_{\rm ADM}$ (solid lines) for each $\lambda$. Thus, Fig.~\ref{fig:E_rad} shows that the peak \(E_{rad}\) for each $\lambda$ occurs at $b=b^*$. For binaries past this peak, an increase in $b$ decreases \(E_{rad}\). Within the $\lambda = 0.6$ cases, there are two BBHs that exhibit zoom-whirl behavior ($b^* < b < b_{\rm scat}$). Among these two, we find that the BBH with the larger $b$, and thus longer time between successive GW bursts,  radiates less energy. BBHs with $b > b_{\rm scat}$ radiate even less energy. 

In Fig.~\ref{fig:J_rad} we show the angular momentum radiated (\(J_{rad}\)) by each binary as a function of $\lambda$ and $b/M_{\rm ADM}$. 
For each $\lambda$, there exists a value of $b$ for which binaries radiate at least $72\%$ of the ADM angular momentum. We find that at low $b$, the initial $\lambda$ has minimal effect. Only as $b/M_{\rm ADM}$ increases, does \(J_{rad}\) show variation with $\lambda$. As $b/M_{\rm ADM}$ approaches $b^*$ and $b_{\rm scat}$, the variation in \(J_{rad}\) shows that charge matters in this regime.

Unlike the maximum \(E_{rad}\), the maximum \(J_{rad}\) is insensitive to $\lambda$ to within the measurement error: the maximum $J_{rad}/J_{\rm ADM}$ for each $\lambda$ lies between $0.72 - 0.73$. The portion of angular momentum radiated via EMWs changes noticeably  across initial $\lambda$, with larger $\lambda$ radiating less angular momentum via GWs and more via EMWs. At $b/M_{\rm ADM} = 1.8$, the angular momentum radiated via EMWs has the anticipated $\lambda^2$ dependence.

While the maximum \(E_{rad}\) appears to coincide with the immediate merger threshold, the maximum \(J_{rad}\) occurs between $b^*$ and $b_{\rm scat}$. Moreover, for  $b^*<b<b_{\rm scat}$, \(J_{rad}\) decreases as $b$ approaches $b_{\rm scat}$. This coincides with the time between the multiple GW bursts increasing. Binaries that scatter radiate even less angular momentum. 

A comparison of Figs.~\ref{fig:E_rad}-\ref{fig:J_rad} with Fig.~\ref{fig:extr} shows that the most extremal merger remnants occur at much smaller impact parameters than the maximum radiated energies and angular momenta. When comparing the plots, note that Fig.~\ref{fig:extr} only includes binaries that merged and so has no data points past the scattering thresholds plotted in Figs.~\ref{fig:E_rad} and \ref{fig:J_rad}.

Lastly, the radiated quantities provide further support for the universality arising from $M_{\rm irr}$. In Fig.~\ref{fig:Erad_combo} (Fig.~\ref{fig:Jrad_combo}) we plot \(E_{rad}\) (\(J_{rad}\))
for each BBH in our study, plotted vs $b/M_{\rm irr}$ (left panel) and vs $a/M_{\rm irr}$ (right panel). The plots demonstrate the alignment (to within determination error) of the $E_{rad}/M_{\rm ADM}$ and $J_{rad}/J_{\rm ADM}$ peaks independently of $\lambda$. For both $E_{rad}$ and $J_{rad}$, the $a/M_{\rm irr}$ parameter aligns the bulk of the curves better than the $b/M_{\rm irr}$ parameter. Figure~\ref{fig:Erad_combo} also shows that at fixed $a/M_{\rm irr}$ or $b/M_{\rm irr}$, larger $\lambda$ have smaller $E_{rad}$; although the energy in EMWs goes up with increasing $\lambda$, it does not compensate for the corresponding decrease in GW energy. 

\subsection{\label{sec:error}Convergence and Error Estimates}

In this section we report the results of our convergence study and estimate errors for several quantities. We choose one of our most challenging cases to conduct this study: a BBH with $\lambda = 0.6$ and $b/M_{\rm ADM} = 2.70$. We choose this case because it yields a nearly-extremal merger remnant with $\Upsilon_f \geq 0.96$, and because BBHs with larger $\lambda$ are more challenging to simulate --- the smaller BH apparent horizon is more difficult to resolve. Therefore, we expect that errors arising from this case are representative for cases with smaller values of $\lambda$. 

In addition to our baseline resolution run, we perform 4 additional simulations of this case. The set of resolutions consists of $h/1.5$, $h/1.25$, $h$, $1.\overline{33} h $, and $ 2 h$, where $h$ denotes the baseline grid spacing of our simulations. We find that our results saturate for resolutions $\Delta x \leq h$, indicating that our baseline resolution has reached the accuracy that can be achieved with these calculations. Thus, to estimate convergence and errors, we adopt the simulations with $\Delta x \geq h$, i.e., $\Delta x \in \{h, 1.\overline{33} h, 2h\}$. 

\begin{figure}
\includegraphics[width=0.46\textwidth]{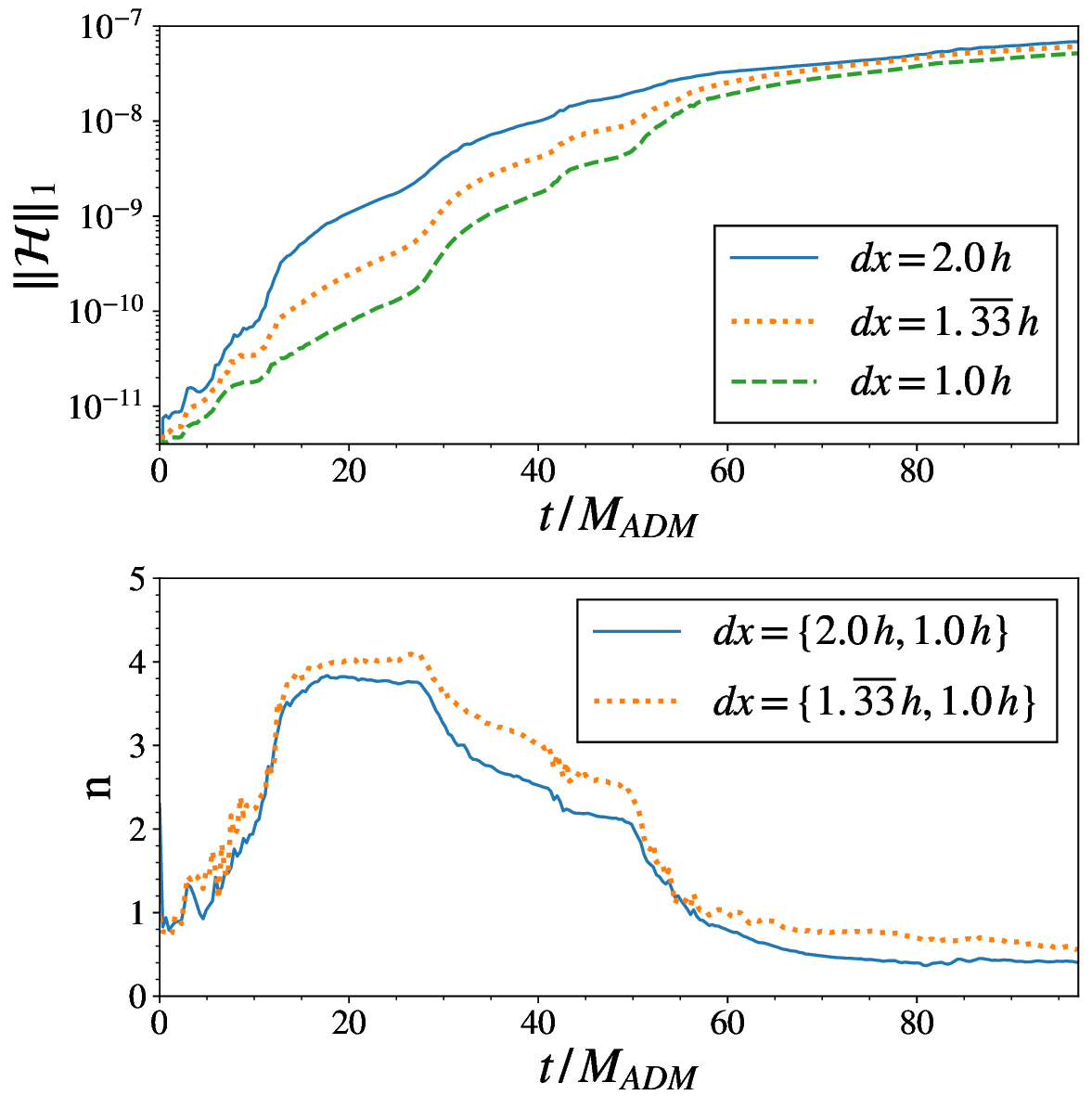}
\caption{\label{fig:ham_constr} Top panel: the $L^1$ norm of the Hamiltonian constraint $\Vert \mathcal{H} \Vert_1$ plotted against time for the three resolutions of our convergence study $\Delta x \in \{ 2.0h, 1.\overline{33} h, 1.0 h\}$. Bottom panel: the order of convergence to zero $n$ of $\Vert \mathcal{H} \Vert_1$ for the two resolution ratios $\Delta x = \{2.0h, 1.0h\}$ and $\Delta x = \{1.\overline{33}h, 1.0h\}$, also plotted against time. The linestyles denote the resolutions used, as described by the legends.}
\end{figure}

In the top panel of Fig.~\ref{fig:ham_constr}, 
we plot the $L_1$ norm of the Hamiltonian constraint ($\Vert \mathcal{H} \Vert_1$) vs time for the three resolutions. The plot demonstrates that the higher the resolution, the lower the Hamiltonian constraint violations become. The bottom panel shows the corresponding order of convergence to zero $n$. $\Vert \mathcal{H} \Vert_1$ exhibits $n \approx 3.8$ for a period prior to merger and drops to $n\approx 1$ after merger. 
We also analyzed the $L_1$ norm of the momentum constraints and find that these converge to zero with increasing resolution, too. In particular, we find $n \approx 2.25 - 2.5$ for the same period that $\Vert \mathcal{H} \Vert_1$ exhibits $n\approx 3.8$.

\begin{table}
\caption{\label{tab:conv_err}
Order of self-convergence $n$ and relative truncation error $E_T$ for the radiated quantities with robust self-convergence, calculated with resolution factors $\Delta x \in \{2.0h, 1.\overline{33}h, 1.0h\}$. The error estimates on $n$ are calculated by varying the lower integration bound by $\pm 10 M_p$. }
\begin{ruledtabular}
\begin{tabular}{ccccc}
  &  $E_{rad}^{GW}$ & $J_{rad}^{EMW}$\\ 
 \hline 
$E_{T} \, (\%)$ &$0.2$ & $1.0$  \\
 $n$  & $3.255 \pm 0.025$ & $3.35 \pm 0.35$ \\
 \end{tabular}
\end{ruledtabular}
\end{table}

The energy radiated via gravitational waves $E_{GW}$ and the angular momentum radiated via electromagnetic waves $J_{EMW}$ display robust convergence. Table~\ref{tab:conv_err}  lists the order of self-convergence $n$ and truncation error $E_T$, computed via Richardson extrapolation for these quantities. The self-convergence of the energy radiated via electromagnetic waves $E_{EMW}$, and the angular momentum radiated via gravitational waves $J_{GW}$ is sensitive to integration bounds. Therefore, we report fractional differences among the different resolutions for fixed integration bounds as rough errors (see Table~\ref{tab:rad_err}).
\begin{table}
\caption{\label{tab:rad_err}
Fractional differences from the reported results in the angular momentum radiated via GWs $J_{rad}^{GW}$ and the energy radiated via EMWs $E_{rad}^{EMW}$ for the two resolutions $\Delta x = 2\, h, 1.\overline{33}\, h$.} 
\begin{ruledtabular}
\begin{tabular}{ccccc}
  &  $\Delta J_{rad}^{GW}(\%)$ & $\Delta E_{rad}^{EMW}(\%)$\\ 
 \hline 
$2\, h$  &$1.7$ & $1.1$  \\
$1.\overline{33} \, h$  & $0.2$ & $0.3$ \\
 \end{tabular}
\end{ruledtabular}
\end{table}

The merger remnant properties do not demonstrate self-convergence. However, the values show minimal variation with resolution. This suggests that either our resolutions have reached the accuracy possible with these simulations, or that the truncation error from the surface integration necessary to compute quasilocal quantities exceeds the error of the simulations. We did not change the resolution in these surface integrations, so we cannot be sure if this is the reason why we did not observe self-convergence in quasilocal quantities of the remnant.

We estimate rough errors in the  remnant properties by calculating the relative difference between the properties produced with $\Delta x = \{2h, 1.\overline{33}h\}$ and the properties produced with $\Delta x = h$ (the reported results). The remnant properties of the simulation with $\Delta x = 2h$ are marked outliers, and their deviations from the reported results could be an overestimation of the error. Likewise, the deviations from the results of the simulation with $\Delta x = 1.\overline{33}h$  could represent an underestimation of the error. We list these errors in Table \ref{tab:resvals_rem}.
\begin{table}
\caption{\label{tab:resvals_rem}
The relative difference of the merger remnant properties calculated with resolution factors $2 \, h$ and $1.\overline{33} \, h$ from the reported results calculated with resolution factor $h$. We list remnant properties mass $M_f$, spin angular momentum $J_f$, dimensionless spin $j_f$, and extremality $\Upsilon_f$.}
\begin{ruledtabular}
\begin{tabular}{ccccc}
                       & $\Delta M_f (\%)$ & $\Delta J_f (\%)$ & $\Delta j_f (\%)$ & $\Delta \Upsilon_f (\%)$\\ 
\hline 
$2 \, h$               & $+ 0.18$         & $+1.28$          & $+0.91$           & $+0.66$ \\
$1.\overline{33} \, h$ & $- 0.04$         & $-0.29$          & $-0.21$           & $-0.15$ \\ 
 \end{tabular}
\end{ruledtabular}
\end{table}

Lastly, we estimate the potential variation due to our initial data interpolation method. After the initial data are generated by \texttt{TwoChargedPunctures}, they must be mapped to the Cartesian grid used by \texttt{Carpet}. \texttt{TwoChargedPunctures} has two methods available for this interpolation: (1) a Taylor interpolation about the nearest collocation point or (2) a spectral interpolation. In this work we used the Taylor interpolant, which is faster but may be less accurate. To confirm that this choice does not influence our results, we performed a simulation of the same case used in the convergence study, adopting our baseline resolution but using the spectral interpolation. The relative variation due to the change in the initial data (ID) interpolation method $E_{ID}$ is listed in Table \ref{tab:error_rad}
\begin{table}
\caption{\label{tab:error_rad}
Initial data interpolation method errors $E_{ID}$ for the radiated quantities, listed as percentages, and calculated for a case with the highest charge-to-mass ratio we explored, $\lambda = 0.6$, and $b/M_{\rm ADM} = 2.70$.}
\begin{ruledtabular}
\begin{tabular}{ccccc}
  &  $E_{rad}^{GW}$ & $E_{rad}^{EMW}$ & $J_{rad}^{GW}$ & $J_{rad}^{EMW}$\\ 
 \hline 
$E_{ID}$ ($\%$) & $0.08$ & $0.1$ & $0.1$ & $0.07$ \\
 \end{tabular}
\end{ruledtabular}
\end{table}
for the radiated quantities and Table \ref{tab:error_rem}
\begin{table}
\caption{\label{tab:error_rem}
Initial data interpolation method errors $E_{ID}$ for merger remnant properties, listed as percentages, and calculated for $\lambda = 0.6$, $b/M_{\rm ADM} = 2.70$.}
\begin{ruledtabular}
\begin{tabular}{ccccc}
  &  $M_f$ & $J_f$ & $j_f$ & $\Upsilon_f$\\ 
 \hline 
 $E_{ID}$ ($\%$) & $0.02$ & $0.04$ & $0.003$ & $0.003$\\
 \end{tabular}
\end{ruledtabular}
\end{table} 
for the merger remnant properties. None of the variations caused by the change in ID interpolation method exceed the other error estimates.
 
\section{\label{sec:Conclusions}Conclusions}

High-energy black hole collisions are testing grounds for fundamental physics. In particular, these interactions allow us to investigate general relativity coupled to a U(1) gauge theory (such as electromagnetism). In this work, we study high-energy collisions near the scattering threshold of equal-mass black holes endowed with the same charge and fixed initial linear  momenta, which correspond to an almost fixed initial Lorentz factor of 1.52. We find remnants that exceed the extremality of remnants generated by initially nonspinning and uncharged BHs to date \cite{sperhake_cross_2009}; our most extremal remnant has a Kerr-Newman extremality parameter $\Upsilon_f = 0.97$ and was generated by a binary with $\lambda = 0.6$. We observe binaries with $\lambda \in \{0.0, 0.1, 0.4, 0.6\}$ that can radiate $28 - 31\%$ of the ADM mass and $72 - 73\%$ of their ADM angular momentum. We note that binaries with higher initial charge-to-mass ratios radiate less energy overall, despite radiating more electromagnetic energy. The initial charge-to-mass ratio has an almost negligible effect on the total angular momentum radiated, but increases the proportion of angular momentum radiated via electromagnetic waves to that radiated via gravitational waves. 

We find that all effects present in uncharged binaries occur at lower $b/M_{\rm ADM}$ in charged binaries. This shift produces strong variation in the dynamics of the binaries and merger remnant properties across $\lambda$ when $b/M_{\rm ADM} > 2.85$. The values of the scattering and immediate merger impact parameter thresholds decrease with increasing charge-to-mass ratio when normalized by $M_{\rm ADM}$, in agreement with the results of Paper I, and the values of $b/M_{\rm ADM}$ that produce the most extremal merger remnant, maximum radiated energy, and maximum radiated angular momentum all decrease as well with increasing initial charge-to-mass ratio. We conclude that charge leaves imprints on key quantities for our moderate Lorentz boost as $b/M_{\rm ADM}$ approaches $b^*/M_{\rm ADM}$, the threshold of immediate merger, and $b_{\rm scat}/M_{\rm ADM}$, the scattering threshold. At $b/M_{\rm ADM} \sim 1.8$, we find that charge has little effect on the radiated quantities and merger remnant properties, indicating that the charge dependence trends to zero as $b$ decreases well below $b^*$. 

As in Paper I, we find that when we normalize the impact parameter by the sum of the BH irreducible masses, the scattering and immediate merger threshold impact parameters become universal, i.e., charge-independent for our fixed initial coordinate separations.  

We observe that the maximum radiated angular momentum and maximum radiated energy occur at the same value of $b/M_{\rm irr}$ across $\lambda$ at a fixed initial coordinate separation. While $b/M_{\rm irr}$ is also almost universal (to within our errors) for the impact parameter of the most extremal merger remnant for each $\lambda$, we find that the specific angular momentum parameter $a = J_{\rm ADM} / M_{\rm ADM}$, normalized by $M_{\rm irr}$, does a better job at providing universality. Note that $a$ has units of impact parameter. All normalized impact parameters that provide universality in some form across $\lambda$ require normalization by $M_{\rm irr}$. These results demonstrate the importance of the areal radius of a BH, which is proportional $M_{\rm irr}$, in setting the fundamental length scale of the problem. In other words, for horizon scale interactions near the scattering threshold, it should be the relative size of the areal radius to the impact parameter that determines the outcome.

The discovered universality with $M_{\rm irr}$ needs to be tested across larger values of $\lambda$ and BHs with spin, as well as unequal mass binaries, and larger Lorentz factors. These will be the topics of an upcoming paper of ours~\cite{MSmith_paper_III}.

\begin{acknowledgments}
We are grateful to the developers and maintainers of the open-source codes that we used. \texttt{Kuibit}~\cite{bozzola_kuibit_2021}, used in our analysis, uses \texttt{NumPy} \cite{harris_array_2020}, \texttt{scipy} \cite{virtanen_scipy_2020}, and \texttt{h5py} \cite{[][{, http://www.h5py.org/.}]noauthor_hdf5_nodate}. \texttt{Matplotlib} \cite{hunter_matplotlib_2007} was used to generate our figures. We thank Vikram Manikantan for his feedback on figures displayed in this work, and Maria Mutz for useful discussions and feedback on the manuscript. This work was in part supported by NSF Grant PHY-2145421 and 
NASA grant 80NSSC24K0771NASA. This work was supported by Advanced Cyberinfrastructure Coordination Ecosystem: Services \& Support (ACCESS) 
allocation TG-PHY190020 and Frontera allocation PHY23009. ACCESS is funded by NSF awards No.~2138259, No.~2138286, No.~2138307, No.~2137603 and No.~2138296 under the Office of Advanced Cyberinfrastructure. The simulations were performed on \texttt{Stampede2} and \texttt{Frontera}, funded by NSF awards No.~1540931 and No.~1818253, respectively, at the Texas Advanced Computing Center (TACC). 

\end{acknowledgments}

\appendix

\begin{table*}[b]
\caption{\label{table:sp_ID} Initial data parameters for the simulations in this work. All binaries have initial linear momentum $\left|P \right| =  0.57236$ and an initial separation $d/M_p = 94.85$. $M_p=1.0$ is the sum of the target quasilocal gravitational masses of the punctures; $M_{\rm ADM}$ is the ADM mass of the binary; $M_{\rm irr}$, $M$, and $Q$ are the sums of the initial irreducible masses, gravitational masses, and charges of the binary's BHs, respectively. A $\checkmark$ in the first column indicates that the binary has $n>1$ encounters prior to merger, i.e., the binary exhibits zoom whirl behavior. $\overline{\lambda}$ is the initial charge-to-mass ratio of the binary, $\overline{\lambda} \equiv Q / M$, computed via the isolated horizon formalism and $\lambda$ is the target initial charge-to-mass ratio. $b/M_p$ and $d/M_p$ are the values input into \texttt{TwoChargedPunctures}. $a \equiv J_{\rm ADM}/M_{\rm ADM}$, where $J_{\rm ADM}$ is the initial angular momentum of the binary.}
\begin{ruledtabular}
\begin{tabular}{ccccccccccccc}
& ZW & $\lambda$ & $\overline{\lambda}$ & $M_{\rm ADM}$& $M$ & $M_{\rm irr}$ & $b/M_p$ & $b/M_{\rm ADM}$ & $b/M_{\rm irr}$ & $a/M_{\rm irr}$ & $d/M_{\rm ADM}$ & \\
\colrule
&  & 0.0 & 0.000 & 1.526 & 1.001 & 1.001 & 2.736 & 1.793 & 2.733 & 1.024 & 62.172 \\
&  & 0.0 & 0.000 & 1.526 & 1.001 & 1.001 & 4.165 & 2.730 & 4.161 & 1.559 & 62.172 \\
&  & 0.0 & 0.000 & 1.526 & 1.001 & 1.001 & 4.560 & 2.989 & 4.556 & 1.707 & 62.172 \\
&  & 0.0 & 0.000 & 1.526 & 1.001 & 1.001 & 4.584 & 3.005 & 4.580 & 1.716 & 62.172 \\
&  & 0.0 & 0.000 & 1.526 & 1.001 & 1.001 & 4.607 & 3.020 & 4.603 & 1.724 & 62.172 \\
&  & 0.0 & 0.000 & 1.526 & 1.001 & 1.001 & 4.621 & 3.029 & 4.616 & 1.729 & 62.172 \\
&  & 0.0 & 0.000 & 1.526 & 1.001 & 1.001 & 4.635 & 3.038 & 4.630 & 1.735 & 62.172 \\
&  & 0.0 & 0.000 & 1.526 & 1.001 & 1.001 & 4.940 & 3.238 & 4.935 & 1.849 & 62.172 \\
&  & 0.0 & 0.000 & 1.526 & 1.001 & 1.001 & 5.100 & 3.343 & 5.095 & 1.909 & 62.172 \\
&  & 0.1 & 0.100 & 1.526 & 1.001 & 0.999 & 2.736 & 1.793 & 2.740 & 1.026 & 62.151 \\
&  & 0.1 & 0.100 & 1.526 & 1.001 & 0.999 & 3.957 & 2.593 & 3.963 & 1.484 & 62.151 \\
&  & 0.1 & 0.100 & 1.526 & 1.001 & 0.999 & 4.165 & 2.729 & 4.171 & 1.562 & 62.151 \\
&  & 0.1 & 0.100 & 1.526 & 1.001 & 0.999 & 4.393 & 2.878 & 4.399 & 1.647 & 62.151 \\
&  & 0.1 & 0.100 & 1.526 & 1.001 & 0.999 & 4.507 & 2.953 & 4.514 & 1.690 & 62.151 \\
&  & 0.1 & 0.100 & 1.526 & 1.001 & 0.999 & 4.550 & 2.982 & 4.557 & 1.706 & 62.151 \\
&  & 0.1 & 0.100 & 1.526 & 1.001 & 0.999 & 4.564 & 2.991 & 4.571 & 1.712 & 62.151 \\
&  & 0.1 & 0.100 & 1.526 & 1.001 & 0.999 & 4.577 & 3.000 & 4.584 & 1.717 & 62.151 \\
&  & 0.1 & 0.100 & 1.526 & 1.001 & 0.999 & 4.590 & 3.008 & 4.597 & 1.721 & 62.151 \\
&  & 0.1 & 0.100 & 1.526 & 1.001 & 0.999 & 4.600 & 3.014 & 4.607 & 1.725 & 62.151 \\
&  & 0.1 & 0.100 & 1.526 & 1.001 & 0.999 & 4.621 & 3.028 & 4.628 & 1.733 & 62.151 \\
&  & 0.1 & 0.100 & 1.526 & 1.001 & 0.999 & 4.900 & 3.211 & 4.907 & 1.838 & 62.151 \\
&  & 0.1 & 0.100 & 1.526 & 1.001 & 0.999 & 4.940 & 3.237 & 4.947 & 1.853 & 62.151 \\
&$\checkmark$ & 0.1 & 0.100 & 1.526 & 1.001 & 0.999 & 4.980 & 3.263 & 4.987 & 1.868 & 62.151 \\
&  & 0.1 & 0.100 & 1.526 & 1.001 & 0.999 & 5.054 & 3.312 & 5.062 & 1.895 & 62.151 \\
&  & 0.4 & 0.399 & 1.534 & 1.002 & 0.960 & 2.736 & 1.784 & 2.849 & 1.062 & 61.834 \\
&  & 0.4 & 0.399 & 1.534 & 1.002 & 0.960 & 3.957 & 2.579 & 4.120 & 1.535 & 61.834 \\
&  & 0.4 & 0.399 & 1.534 & 1.002 & 0.960 & 4.365 & 2.846 & 4.546 & 1.694 & 61.834 \\
&  & 0.4 & 0.399 & 1.534 & 1.002 & 0.960 & 4.380 & 2.855 & 4.561 & 1.699 & 61.834 \\
&  & 0.4 & 0.399 & 1.534 & 1.002 & 0.960 & 4.393 & 2.864 & 4.575 & 1.704 & 61.834 \\
&  & 0.4 & 0.399 & 1.534 & 1.002 & 0.960 & 4.406 & 2.872 & 4.588 & 1.709 & 61.834 \\
&  & 0.4 & 0.399 & 1.534 & 1.002 & 0.960 & 4.435 & 2.891 & 4.619 & 1.721 & 61.834 \\
&  & 0.4 & 0.399 & 1.534 & 1.002 & 0.960 & 4.460 & 2.908 & 4.645 & 1.730 & 61.834 \\
&  & 0.4 & 0.399 & 1.534 & 1.002 & 0.960 & 4.500 & 2.934 & 4.686 & 1.746 & 61.834 \\
&  & 0.4 & 0.399 & 1.534 & 1.002 & 0.960 & 4.621 & 3.012 & 4.812 & 1.793 & 61.834 \\
&  & 0.4 & 0.399 & 1.534 & 1.002 & 0.960 & 4.735 & 3.087 & 4.931 & 1.837 & 61.834 \\
&$\checkmark$ & 0.4 & 0.399 & 1.534 & 1.002 & 0.960 & 4.780 & 3.116 & 4.978 & 1.855 & 61.834 \\
&  & 0.4 & 0.399 & 1.534 & 1.002 & 0.960 & 4.830 & 3.149 & 5.030 & 1.874 & 61.834 \\
&  & 0.4 & 0.399 & 1.534 & 1.002 & 0.960 & 4.940 & 3.221 & 5.144 & 1.917 & 61.834 \\
&  & 0.6 & 0.598 & 1.545 & 1.003 & 0.904 & 2.736 & 1.770 & 3.028 & 1.120 & 61.377 \\
&  & 0.6 & 0.598 & 1.545 & 1.003 & 0.904 & 3.957 & 2.560 & 4.378 & 1.619 & 61.377 \\
&  & 0.6 & 0.598 & 1.545 & 1.003 & 0.904 & 4.145 & 2.682 & 4.587 & 1.696 & 61.377 \\
&  & 0.6 & 0.598 & 1.545 & 1.003 & 0.904 & 4.165 & 2.695 & 4.609 & 1.704 & 61.377 \\
&  & 0.6 & 0.598 & 1.545 & 1.003 & 0.904 & 4.195 & 2.715 & 4.642 & 1.717 & 61.377 \\
&  & 0.6 & 0.598 & 1.545 & 1.003 & 0.904 & 4.222 & 2.732 & 4.672 & 1.728 & 61.377 \\
&  & 0.6 & 0.598 & 1.545 & 1.003 & 0.904 & 4.279 & 2.769 & 4.735 & 1.751 & 61.377 \\
&  & 0.6 & 0.598 & 1.545 & 1.003 & 0.904 & 4.393 & 2.843 & 4.861 & 1.798 & 61.377 \\
&  & 0.6 & 0.598 & 1.545 & 1.003 & 0.904 & 4.470 & 2.893 & 4.946 & 1.829 & 61.377 \\
&$\checkmark$ & 0.6 & 0.598 & 1.545 & 1.003 & 0.904 & 4.507 & 2.916 & 4.987 & 1.844 & 61.377 \\
&$\checkmark$ & 0.6 & 0.598 & 1.545 & 1.003 & 0.904 & 4.540 & 2.938 & 5.024 & 1.858 & 61.377 \\
&  & 0.6 & 0.598 & 1.545 & 1.003 & 0.904 & 4.621 & 2.990 & 5.113 & 1.891 & 61.377
\end{tabular}
\end{ruledtabular}
\end{table*}

\section{\label{sec:list_id}List of Initial Data}

\begin{figure*}
\includegraphics[width=0.95\textwidth]{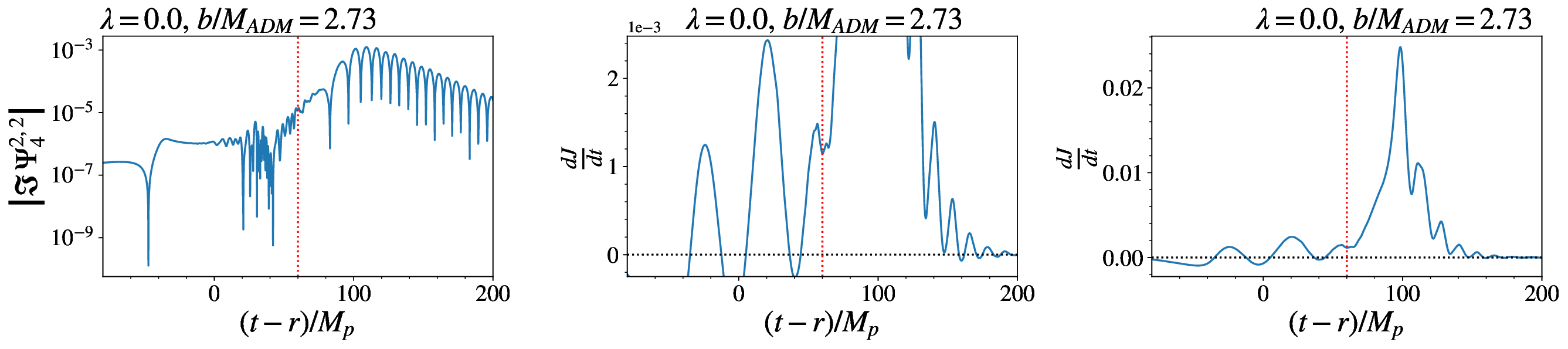}
\caption{\label{fig:Jbound_ex1} The absolute value of the imaginary part of the $2,2$ mode of $\Psi_4$ (left), a zoom-in on the angular momentum radiated as a function of time $\frac{dJ}{dt}$ (center), and the full angular momentum radiated curve (right) for a binary with $\lambda = 0.0$ and $b/M_{\rm ADM} = 2.73$. The vertical red dashed line is the lower bound from which we integrate $\frac{dJ}{dt}$. The left panel uses a log-scale along the y-axis. We display the imaginary component of $\Psi_4^{2,2}$ because it is used in the calculation of $\frac{dJ}{dt}$. }
\end{figure*}

The precise properties specifying the initial data for our 51 simulations are presented in Table~\ref{table:sp_ID}. All BHs have initial linear momentum $\left|P \right| =  0.57236$, which corresponds to a target BH mass of $0.5$ and initial boost $\gamma = 1.520$. The gravitational quasilocal BH mass varies slightly, and the initial boost compensates to produce a constant $\left| P \right|$. 

\section{\label{sec:Rad_cutoffs} Identification of Integration Bounds for Radiated Angular Momentum Calculation}


As described in Sec.~\ref{subsec:diag}, we use Eq.~19c of \cite{bozzola_numerical-relativity_2021} to calculate the angular momentum radiated by each binary via GWs. We integrate in the time domain, up to and including $\ell = 6$ spin weighted spherical harmonics with weight $s=-2$. We excise unphysical radiation by implementing upper and lower bounds over which to integrate the angular momentum radiated per unit time $\frac{dJ}{dt}$. The lower bound is used to remove the junk radiation, and its selection is not always straight-forward. Here we give a more in-depth explanation of this process. 

The first step in this calculation is identifying the upper integration bound, which coincides with when noise begins to dominate the $2,2$ mode of the Weyl tensor $\Psi_4$. We then calculate $\frac{dJ}{dt}$ without any cropping or windowing of the original $\Psi_4$ data and subtract off the integration constant at the upper bound for each integration. The final step is identifying the lower bound and integrating $\frac{dJ}{dt}$ from the lower to the upper bound.    

The identification of the lower bound is the most challenging step of this process. Here we give an example. The left panel of Fig.~\ref{fig:Jbound_ex1}
displays the absolute value of the imaginary component of the $2,2$ mode of $\Psi_4$ plotted on a logscale for one of our binaries ($\lambda = 0.0$, $b/M_{\rm ADM} = 2.73$).
The center and right panels display the $\frac{dJ}{dt}$ for the binary, with the center panel a zoomed-in version of the right panel. The $\frac{dJ}{dt}$ curve has a peak around $(t - r)/M_p = 100$ that corresponds to the merger of the BHs. 
While the luminosity curves have only one peak corresponding to junk radiation, the $\frac{dJ}{dt}$ curves have multiple, making the determination of the lower bound more challenging.
Inspection of $\left| \Im{ (\, \Psi_4^{2,2})} \right|$ tells us that the junk radiation persists until at least $(t - r)/M_p = 50$ and that the merger occurs around $(t - r)/M_p = 100$. We therefore select the lower bound as the local minimum immediately following  $(t - r)/M_p = 50$. This is also the  local minimum immediately preceding the peak in $\frac{dJ}{dt}$ corresponding to merger. We repeat this process for all of our binaries, identifying the peaks in $\frac{dJ}{dt}$ that are generated by junk radiation and those that are generated by the merger. The selection of the integration bound has moderate affect on the energy radiated via gravitational waves $E^{GW}_{rad}$ and a more significant effect on the angular momentum $J^{GW}_{rad}$: for the binary used in our convergence study with $\lambda = 0.6$, $b/M_{\rm ADM} = 2.70$, $E_{rad}^{GW}$ varied by $0.9\%$ and $J_{rad}^{GW}$ varied by $5.3\%$ when the lower integration bound was moved by $\pm 10 M_p$. 

\section{Explanation of peak locations for different remnant quantities \label{app:peaks_exp}}

In this appendix we explain why $j_f$ and $\Upsilon_f$ for given value of $\lambda$ generally peak at different values of $b$. We also explain why $J_f$ and $j_f$ peak at different values of $b$.

As seen in Fig.~\ref{fig:extr}, for $\lambda = 0.6$, the peak in the $\Upsilon_f$ curve occurs at a larger value of $b/M_{\rm ADM}$ than the peak in the $j_f$ curve. 
There is a straightforward mathematical explanation for this effect: For a given $\lambda$, the change of $\Upsilon_f$ with $b$ can be derived from Eq.~\eqref{eq:chi_formula} and is given by 
\begin{eqnarray}
\frac{d \, \Upsilon_f}{db} 
= \frac{1}{\Upsilon_f}\left[j_f \left(\frac{d \, j_f}{db}\right) + \lambda_f \left(\frac{d \, \lambda_f}{db}\right)\right].
\label{eq:chi_peak}
\end{eqnarray}
Equation~\eqref{eq:chi_peak} demonstrates that when $\lambda_f$ is non-zero, the location of the $\Upsilon_f$--$b$ peak (where $\frac{d \, \Upsilon_f}{db} =0$) does not coincide with the $j_f$--$b$ peak (where $\frac{d\, j_f}{db} = 0$), because $\lambda_f$ varies with $b$. Our simulations show that the effect is exaggerated as $\lambda_f$ (and $\lambda$) increases.

Additionally, as shown by Fig.~\ref{fig:J_v_j}, the maximum $J_f$ for fixed $\lambda$ occurs at a different value of $b/M_{\rm ADM}$ than where the maximum $j_f$ occurs. This result is a consequence of the definition of dimensionless spin. The derivative of $j_f$ with respect to $b$ is given by
\begin{eqnarray}
\frac{d \, j_f}{db} 
= \frac{1}{M_f^2}\left[\left(\frac{dJ_f}{db}\right) - 2 \frac{J_f}{M_f}\left(\frac{dM_f}{db}\right)\right].
\label{eq:J_peak}
\end{eqnarray}
The maximum in the $j_f$--$b$ curve (where $\frac{d\, j_f}{db} = 0$) is offset from the maximum of the $J_f$--$b$ curve (where $\frac{d\, J_f}{db} = 0$), because $\frac{dM_f}{db}\neq 0$.

\section{\label{sec:Chi_err}Calculation of $\delta \Upsilon_f^{max}\left(\overline{\lambda}\right)$}

Prior to performing a least squares fit for the $\Upsilon_f^{max}$--$\overline{\lambda}$ curve shown in Fig. \ref{fig:proj_chi}, we first estimate an error for each $\Upsilon^{max}_f$--$\overline{\lambda}$ data point, which we feed into our fit algorithm. This error, $\delta \Upsilon_f^{max} \left(\overline{\lambda}\right)$, is due to the finite sampling in impact parameter space, which does not allow a perfect determination of the $b$ resulting in $\Upsilon_f^{max}$ for a given $\lambda$. For some values of $\lambda$ in our suite, like $\lambda = 0.1$, we have located $\Upsilon_f^{max}$ in the $\Upsilon_f$--$a/M_{\rm irr}$ curve (see Fig. \ref{fig:chi_alpha}) to high precision. Others, like $\lambda = 0.4$, have been determined with lower precision. The error estimate $\delta \Upsilon_f^{max}\left(\overline{\lambda}\right)$ quantifies for the $\Upsilon_f^{max}$--$\overline{\lambda}$ best-fit routine these differences, e.g., that the $\lambda = 0.4$ data point should carry slightly less weight than the $\lambda = 0.1$ data point.  

To compute $\delta \Upsilon_f^{max}\left(\overline{\lambda}\right)$, we repeat the following process for each $\lambda$ in the respective $\Upsilon_f$--$a/M_{\rm irr}$ curve. First, for each $\lambda$ we fit a (different) second-order polynomial to the $3$ data points at the peak in the $\Upsilon_f$--$a/M_{\rm irr}$ curve  using \texttt{NumPy}'s \cite{harris_array_2020} \texttt{polyfit}. We designate the maximum $\Upsilon_f$ predicted by the $\Upsilon_f$--$a/M_{\rm irr}$ fit by $\Upsilon_f^{max, \, projected}$.  The value of $\Upsilon_f^{max, \, projected}$ is compared to the largest value of $\Upsilon_f$ in our dataset for that $\lambda$, i.e., the largest data point, $\Upsilon_f^{max,\, data}$: 
\begin{eqnarray}
\delta \, \Upsilon_f^{max}  = \left|\Upsilon_f^{max, \, data} - \Upsilon_f^{max, \, projected} \right|
\label{eq:deltachi}.
\end{eqnarray}
This process is repeated for each $\overline{\lambda}$, generating $\delta \Upsilon_f^{max}\left(\overline{\lambda}\right)$.



%

\end{document}